\documentclass[10pt,journal,compsoc]{IEEEtran}
\usepackage{amsmath,amssymb,amsfonts}
    \usepackage{graphicx}
    \usepackage[T1]{fontenc} 
    \usepackage[cmintegrals]{newtxmath}
    \usepackage{comment}
    \usepackage{textcomp}
    \usepackage{bm}
    \usepackage{nicematrix}
    \usepackage{tikz}
    \usepackage{tabu}
    \usepackage[lined,boxed,ruled]{algorithm2e}
    \usepackage{multirow}
    \usepackage{booktabs}
    \usepackage{stfloats}
    \usepackage{times}
    \usepackage{epsfig}
    \usepackage{bm}
    \usepackage{color}
    \usepackage{arydshln}
    \usepackage{mathrsfs}
    \usepackage{subcaption}
    \usepackage{tabularx}
\PassOptionsToPackage{hyphens}{url}
\usepackage[letterpaper=true,colorlinks,bookmarks=false]{hyperref}
\usepackage{xcolor}
\usepackage{pifont}
\newcommand{\cmark}{\ding{51}}%
\newcommand{\xmark}{\ding{55}}%
    
\ExplSyntaxOn

\keys_define:nn { MyRule }
  {
    color .tl_set:N = \l__MyRule_color_tl , 
    color .value_required:n = true ,
    width .dim_set:N = \l__MyRule_width_dim ,
    width .value_required:n = true ,
    width .initial:n = \arrayrulewidth ,
    style .value_required:n = true ,
    style .tl_set:N = \l__MyRule_style_tl
  }

\NewDocumentCommand { \MyRule } { O { } D ( ) { } m }
  {
    \exp_args:Nx \__MyRule_i:nnnn 
    { \int_use:c { c@iRow } } { #1 } { #2 } { #3 } 
  }

\cs_new_protected:Nn \__MyRule_i:nnnn
  {
    \tl_gput_right:Nn \g_nicematrix_code_after_tl 
      { \__MyRule_ii:nnnn { #1 } { #2 } { #3 } { #4 } }
    \peek_remove_spaces:n { }
  }

\cs_new_protected:Nn \__MyRule_ii:nnnn
  {
    \group_begin:
    \keys_set:nn { MyRule } { #2 }
    \__MyRule_iii:w #4 \q_stop
    \begin { tikzpicture }
    \pgfsetlinewidth { \l__MyRule_width_dim }
    \tl_if_empty:NF \l__MyRule_color_tl    
      { \pgfsetstrokecolor { \l__MyRule_color_tl } }
    \tl_if_in:nnTF { #3 } { l }
      { \dim_set:Nn \l_tmpa_dim { 1 mm } }
      { \dim_zero:N \l_tmpa_dim }
    \tl_if_in:nnTF { #3 } { r }
      { \dim_set:Nn \l_tmpa_dim { 1 mm } }
      { \dim_zero:N \l_tmpa_dim }
    \tl_if_empty:NF \l__MyRule_line_style_tl
      { \tikzset { every~path/.style = \l__MyRule_style_tl  } }
    \draw ([xshift=\l_tmpa_dim,yshift=-\l__MyRule_width_dim/2] #1 
             -| \int_use:N \l_tmpa_int ) --
          ([xshift=-\l_tmpa_dim,yshift=-\l__MyRule_width_dim/2] #1 
             -| \int_use:N \l_tmpb_int )  ;
    \end { tikzpicture }
    \group_end:
  }

\cs_new_protected:Npn \__MyRule_iii:w #1 - #2 \q_stop
  {
    \int_set:Nn \l_tmpa_int { #1 }
    \int_set:Nn \l_tmpb_int { #2 + 1 }
  }

\ExplSyntaxOff

\ifCLASSOPTIONcompsoc
  \usepackage[nocompress]{cite}

\else
  \usepackage{cite}
  
\fi

\ifCLASSINFOpdf
\else
\fi

\begin{document}
%
\title{Fourier-Net+: Leveraging Band-Limited Representation for Efficient 3D \\ Medical  Image Registration}

\author{Xi Jia, Alexander Thorley, Alberto Gomez, Wenqi Lu, Dipak Kotecha and Jinming Duan
\IEEEcompsocitemizethanks{
\IEEEcompsocthanksitem X. Jia, A. Thorley, and J. Duan are with the School of Computer Science, University of Birmingham, Birmingham, UK.  
\IEEEcompsocthanksitem A. Thorley and A. Gomez are with the Ultromics Ltd, Oxford, UK.
\IEEEcompsocthanksitem W. Lu is with the Department of Computer Science, University of Warwick, UK. 
\IEEEcompsocthanksitem D. Kotecha is with the Institute of Cardiovascular Sciences, University of Birmingham, Birmingham, UK.
\IEEEcompsocthanksitem J. Duan is also with the Alan Turing Institute, London, UK. 
\IEEEcompsocthanksitem The corresponding author is J. Duan (email: j.duan@bham.ac.uk).}


\thanks{Manuscript received xxxx xx, 2022}}

%
%

\markboth{Journal of \LaTeX\ Class Files,~Vol.~14, No.~8, August~2015}%
{Shell \MakeLowercase{\textit{et al.}}: Bare Advanced Demo of IEEEtran.cls for IEEE Computer Society Journals}
 
\IEEEtitleabstractindextext{%
\begin{abstract}
U-Net style networks are commonly utilized in unsupervised image registration to predict dense displacement fields in the full-resolution spatial domain. For high-resolution volumetric image data, this process is however resource-intensive and time-consuming. To tackle this challenge, we first propose Fourier-Net, which replaces the U-Net style network's expansive path with a parameter-free model-driven decoder. This results in fewer network parameters, memory usage, and computational operations. Specifically, instead of directly predicting a full-resolution displacement field in the spatial domain, our Fourier-Net learns a low-dimensional representation of the displacement field in the band-limited Fourier domain. This representation is then decoded by our model-driven decoder to obtain the dense, full-resolution displacement field in the spatial domain. Expanding upon Fourier-Net, we then introduce Fourier-Net+, which takes the band-limited spatial representation of the images as input, instead of their original full-resolution counterparts. This leads to a reduction in the number of convolutional layers in the U-Net style network's contracting path, resulting in a further decrease in network parameters, memory usage, and computational operations. Finally, to enhance the registration performance, we propose a cascaded version of Fourier-Net+. We evaluate our proposed methods on three datasets, including two brain and a cardiac MRI (CMR) datasets, comparing them against various state-of-the-art approaches. Our proposed Fourier-Net and its variants achieve comparable results with these approaches, while exhibiting faster inference speeds with a lower memory footprint and fewer multiply-add operations. For example on the 3D-CMR dataset, our Fourier-Net+ outperforms the current state-of-the-art models, TransMorph and LKU-Net, with improvements of 7.1\% and 7.5\% in terms of Dice, respectively. Additionally, our Fourier-Net+ exhibits exceptional inference speed,  9.05 and 4.42 times faster, while utilizing only 0.35\% and 0.84\% of their multiply-add operations, and 2.07\% and 0.97\% of their memory usage. With such small computational cost, our Fourier-Net+ enables the training of large-scale 3D registration on low-VRAM GPUs efficiently. Our code is publicly available at \url{https://github.com/xi-jia/Fourier-Net}.

\end{abstract}

}

\maketitle

\IEEEdisplaynontitleabstractindextext

%
\IEEEpeerreviewmaketitle

\ifCLASSOPTIONcompsoc
\IEEEraisesectionheading{\section{Introduction}\label{sec:introduction}}
\else
\section{Introduction}
\label{sec:introduction}
\fi

\IEEEPARstart{M}{edical} image registration plays a critical role in many medical image analysis applications, such as population modeling, longitudinal studies, and statistical atlases \cite{sotiras2013deformable}. The goal of medical image registration is to learn a spatial deformation that establishes the correspondences between a fixed image and a moving image. Medical image registration can be categorized based on various factors, including data dimensionality (e.g., 2D, 3D, and 3D$+t$), image modality (e.g., CT, MR, and Ultrasound), objects of interest (e.g., brain, lung, and heart), nature of transformation (e.g., rigid, affine, and deformable), registration basis (e.g., landmarks, image features, and voxel intensities), and optimization techniques (e.g., gradient descent or ADMM). For a more comprehensive taxonomy of medical image registration, we suggest that the reader refers to the surveys presented in \cite{maintz1998survey} and the updated version in \cite{viergever2016survey}. In this work, we concentrate on unsupervised, intensity-based, mono-modality, deformable image registration.

Intensity-based deformable image registration has been an active field for the past two decades. Given a moving image and a fixed image, the goal is to compute a dense deformation field that minimizes the distance (or maximizes the similarity) between the warped moving image and the fixed image. While the concept of deformable image registration is simple, it remains a challenging problem due to its ill-posed nature. That is, the solution is often not unique \cite{modersitzki2003numerical}. In practice, prior constraints or assumptions over the deformation are imposed before its estimation, including but not limited to smoothness, symmetry, topology preservation, and diffeomorphisms \cite{sotiras2013deformable}. These prior constraints serve as regularization terms that are imposed on the deformation, in combination with data distance (or similarity) terms between the warped and fixed images to construct the optimization objective of registration methods. In general, deformable image registration for medical imaging can be categorized as iterative optimization-based registration approaches, their deep learning-based counterparts, and a combination of both approaches.




Before the emergence of deep learning, deformable image registration relied heavily on iterative optimization techniques. Popular optimization-based approaches from this paradigm include free-form deformation (FFD) \cite{rueckert1999nonrigid}, large deformation diffeomorphic metric mapping (LDDMM) \cite{beg2005computing}, DARTEL \cite{ashburner2007fast}, Demons\cite{vercauteren2009diffeomorphic}, Elastix \cite{klein2009elastix}, ANTs \cite{avants_ANTS}, NiftyReg \cite{modat2010fast}, Flash \cite{zhang2019fast}, ADMM \cite{thorley2021nesterov,duan2023arbitrary}, etc. Though widely applied and mathematically sound, these approaches require extensive hyper-parameter tuning for each image pair and are computationally expensive thus limiting their applications in real-time and large-scale volumetric image registration.


\begin{figure}[t]
    \centering
    \includegraphics[width=0.48\textwidth]{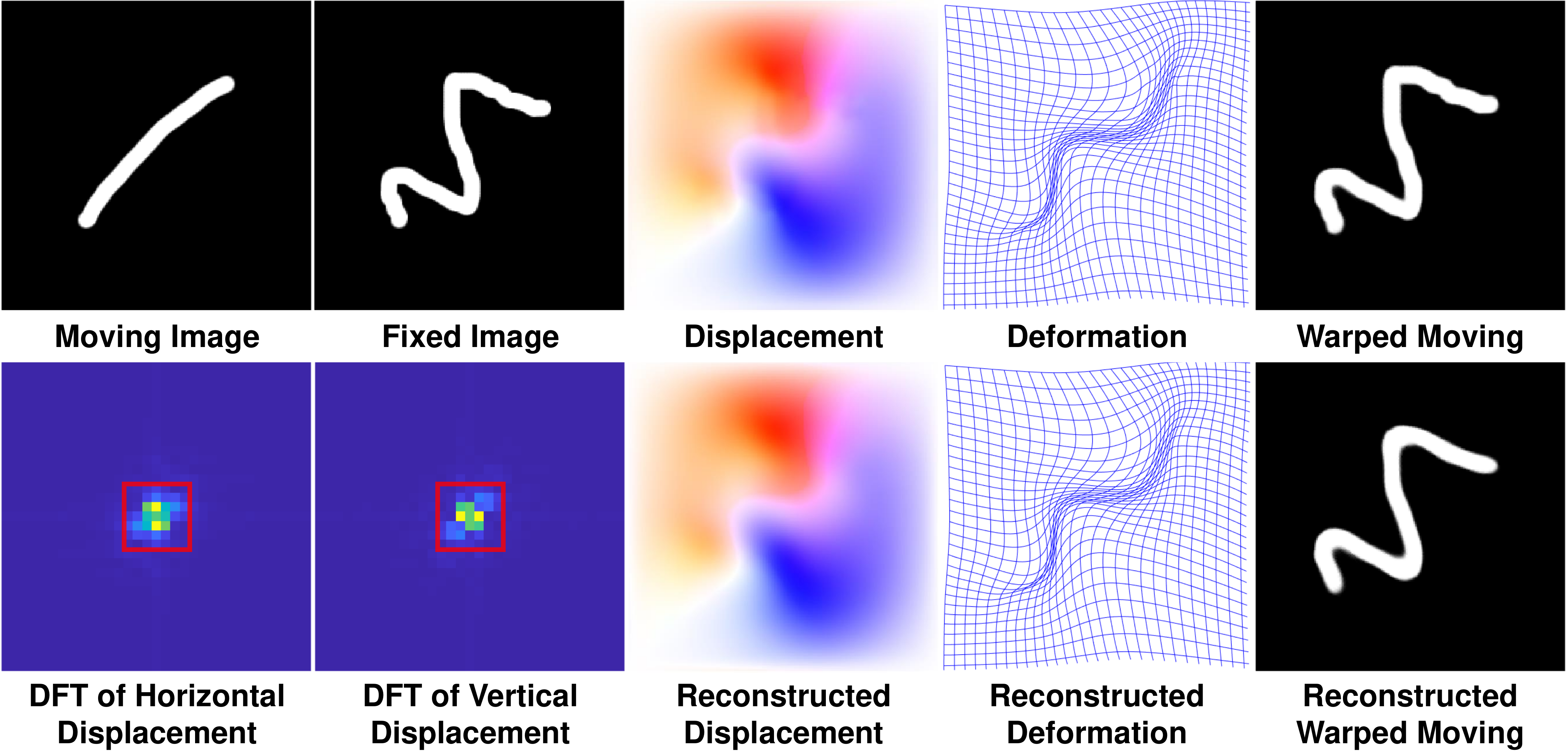}
    \caption{An example showing connections between deformations in the spatial domain and band-limited Fourier domain. From left to right in the 1st row: moving image, fixed image, displacement field, deformation grid, and warped moving image. From left to right in the 2nd row: Discrete Fourier transformation (DFT) of the horizontal displacement field, DFT of the vertical displacement field,  reconstructed displacement field from only the Fourier coefficients within the band-limited region (marked by red rectangles), reconstructed deformation grid, and warped moving image using reconstructed deformation.}
    \label{fig:one}
\end{figure}

Recently, there has been a surge in the use of deep learning-based approaches for medical image registration \cite{hering2022learn2reg}. Their success is largely due to their ability to perform fast inference, and the flexibility to leverage auxiliary information such as anatomical masks as part of the training process. The most effective methods, such as VoxelMorph \cite{balakrishnan2019voxelmorph}, typically employ a U-Net style architecture to estimate dense spatial deformation fields. These methods require only one forward pass during inference, making them orders of magnitude faster than traditional iterative methods. Following the success of VoxelMorph, numerous deep neural networks have been proposed for various registration tasks \cite{zhang2018inverse, hering2019mlvirnet, zhao2019unsupervised, Zhao_2019_ICCV, Mok_2020_CVPR, kim2021cyclemorph, chen2021transmorph, jia2022u}. These networks generally enhance the registration performance through two strategies: cascading U-Net style networks \cite{hering2019mlvirnet, zhao2019unsupervised, Zhao_2019_ICCV} and replacing basic convolutional blocks with more sophisticated alternatives, such as attention-based transformers \cite{chen2021transmorph} and parallel convolutional blocks \cite{jia2022u}. However, these changes increase the number of network parameters and multiply-add operations  (mult-adds), negatively impacting training and inference efficiency.



An alternative approach for image registration is to combine data-driven deep learning models with iterative methods, as proposed by \cite{jia2021learning, blendowski2021weakly, qiu2022embedding}. These methods embed the mathematical structure of minimizing a generic objective model into a neural network. By doing so, the network mapping process inherits prior knowledge from the objective model while maintaining the data efficiency of the iterative methods. As a result, these model-driven networks have the advantages of both communities and have been shown to outperform purely data-driven registration methods. However, despite being faster than traditional iterative optimization-based methods, emulating the iterative optimization process often requires the use of multiple U-Nets and other sophisticated neural layers in these networks (e.g., intensity consistency layer \cite{jia2021learning}), which often result in a slower registration speed when compared to purely network-based methods.

A common characteristic among learning-based registration approaches is the utilization of U-Net style networks. In this paper, we argue that for such styles of registration networks, it may not be necessary to include the entire expansive path of the decoder. Additionally, we suggest that training and inference efficiency can be further improved by learning a low-dimensional representation of the displacement field in the band-limited Fourier domain. Our observations are based on the results shown in Figure~\ref{fig:one}, which demonstrate that a small number of coefficients in the band-limited Fourier domain are sufficient to reconstruct a full-resolution deformation accurately. Inspired by this insight, we propose Fourier-Net, an end-to-end unsupervised registration model that is able to learn such a low-dimensional, band-limited representation of the displacement field. Specifically, by removing several layers in the expansive path of a U-Net style architecture, Fourier-Net outputs only a small patch containing low-frequency coefficients of the displacement field in the Fourier domain. A model-driven decoder then recovers the full-resolution spatial displacement field from these coefficients, using a zero-padding layer that broadcasts complex-valued low-frequency signals into a full-resolution complex-valued map, and an inverse discrete Fourier transform (iDFT) that recovers the spatial displacement field from the map. Both zero-padding and iDFT layers are parameter-free, making our Fourier-Net very fast. We also propose a diffeomorphic variant, termed Diff-Fourier-Net, which learns the band-limited representation of the velocity field and uses the squaring and scaling layers to encourage the output deformation to be diffeomorphic \cite{dalca2018unsupervised}.

\begin{figure}
    \centering
    \includegraphics[width=0.46\textwidth]{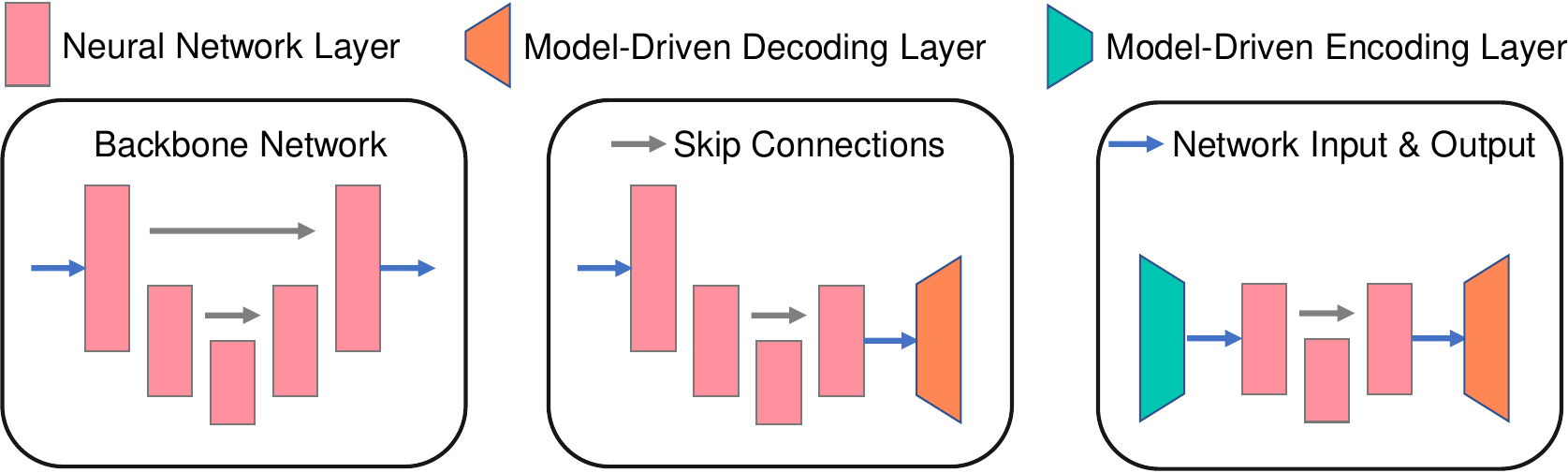}
    \caption{Schematic overview of deep registration models. Left: U-Net style architecture that most models employed to predict the dense deformation field. Middle: Approaches like B-Spline \cite{qiu2021learning}, DeepFlash\cite{wang2020deepflash}, and our Fourier-Net that discard part of the expansive path and use model-driven decoding function/layer instead. Right: The proposed Fourier-Net+, which has a model-driven encoding layer and decoding layer, greatly reducing the parameters and computational costs of the network.}
    \label{fig:SchematicOverview}
\end{figure}

Building on the results of Fourier-Net, we hypothesize that it may be feasible to learn the band-limited displacement field or band-limited velocity field directly from a band-limited representation of the input image pairs, rather than from the original full-resolution image pairs. This has the potential to further reduce the number of convolutional layers in the contracting path in Fourier-Net and further accelerate its registration speed. To this end, we propose Fourier-Net+, which utilizes the same decoder as Fourier-Net but features an improved encoder that aims to reduce computational overhead. As per standard U-Net style architectures, Fourier-Net is designed to take the original full-resolution image pairs as input, with an encoder involving multiple 3D convolutional layers which are computationally expensive operations, particularly in the earlier layers of the model. In contrast, the encoder of Fourier-Net+ includes a model-driven encoding of the original full-resolution image pairs into a low-dimensional band-limited representation of such image pairs, followed by 3D convolutions. This encoder enables 3D convolutions to operate on smaller resolutions, making Fourier-Net+ a much lighter network compared to Fourier-Net or U-Net style backbones. Figure~\ref{fig:SchematicOverview} provides a visual comparison of the architectural differences between U-Net, Fourier-Net, and Fourier-Net+. To enhance the registration performance of Fourier-Net+, we then propose a cascaded version of this network: cascaded Fourier-Net+. Despite the potential increase in computational cost from using multiple cascades, the efficient design of Fourier-Net+ allows the cascaded version of this network to still have fewer mult-adds operations than Fourier-Net in our experiments. We note that preliminary results of Fourier-Net were presented in the conference proceedings \cite{jia2023fourier}. In this paper, we introduce the two extended versions of Fourier-Net, namely Fourier-Net+ and cascaded Fourier-Net+. We also provide extensive experimental results to thoroughly investigate Fourier-Net and its variations.

\section{Related Works}

\subsection{Iterative Optimization-Based Approaches} Iterative methods based on instance-level optimization are prohibitively slow, especially when the images to be registered are of a high-dimensional form. Over the past decades, many works have been proposed to accelerate such methods. For example, rather than directly estimating dense displacement fields, FFD models \cite{rueckert1999nonrigid} were proposed in which a deformation grid over a few control points is interpolated to a dense field using B-splines. 
Computational challenges in diffeomorphic registration are even more pronounced. For this task, Ashburner \cite{ashburner2007fast} introduced a fast algorithm called DARTEL for diffeomorphic registration to accelerate the computation of LDDMM, which integrates deformations through non-stationary velocities over time using the Lagrange transport equation. DARTEL used a stationary velocity field (SVF) representation \cite{arsigny2006log, legouhy2019unbiased} and computed the resulting deformation through scaling and squaring of the SVF. Zhang and Fletcher \cite{zhang2019fast} developed the Fourier-approximated Lie algebras for shooting (Flash) for fast diffeomorphic image registration, where they proposed to speed up the solution of the EPDiff equation used to compute deformations from velocity fields in the band-limited Fourier domain. Hernandez \cite{hernandez2018band} reformulated the Stokes-LDDMM variational problem used in \cite{mang2015inexact} in the domain of band-limited non-stationary vector fields and utilized GPUs to parallelize their methods. Another fast approach for deformable image registration is Demons \cite{vercauteren2009diffeomorphic}, which imposed smoothness on displacement fields by incorporating inexpensive Gaussian convolutions into its iterative process. The diffeomorphisms in Demons were achieved by the greedy composition of the speed vector fields \cite{vercauteren2009diffeomorphic}. Recently, Thorley et al. \cite{thorley2021nesterov} proposed a convex optimization model that used an arbitrary order regularisation term. By combining Nesterov's accelerated gradient descent and the alternating direction method of multipliers (ADMM), this model was shown to register pairs of cardiac 3D MR images within 2s on GPUs. 


\subsection{Deep Registration Approaches}

\textbf{Unsupervised methods:} 
Convolutional neural networks (CNNs) based on unsupervised learning have recently been used to accelerate the speed of registration \cite{qin2018joint, balakrishnan2019voxelmorph,hu2019dual} without the need for ground truth deformations. U-Net style networks, in particular, have been shown to be effective in learning deformations between pairwise images \cite{balakrishnan2019voxelmorph,zhang2018inverse,Mok_2020_CVPR,kim2021cyclemorph,jia2021learning}. The authors of VoxelMorph \cite{balakrishnan2019voxelmorph} demonstrated that a simple U-Net can achieve comparable registration performance to iterative methods, with inference times orders of magnitude faster. Building on the success of VoxelMorph, various extensions have been proposed to improve registration accuracy and address specific challenges in medical image registration.


One such extension used multiple recurrent or cascaded U-Net architectures, where the deformation was iteratively composed by incorporating the output from each cascade \cite{hering2019mlvirnet, zhao2019unsupervised, Zhao_2019_ICCV}. These methods were shown to be particularly effective at handling large deformations in input image pairs by allowing the network to capture both global and local features of the images. Another approach to improving U-Net-based registration models was to augment the U-Net backbone with more representative layers to better capture correspondences between input images. For example, transformer-based methods, such as ViT-V-Net \cite{chen2021vit} and TransMorph \cite{chen2021transmorph}, replaced some of the regular convolutional layers in the U-Net style network with transformer layers, allowing for the modeling of long-range dependencies between voxels, and LKU-Net \cite{jia2022u} employed parallel large-kernel convolutional layers to handle different scales in input images. Although combining U-Nets with more representative layers has been shown to achieve promising registration performance, these methods come with a significantly higher computational cost, leading to slower inference speed. Another alternative to U-Net style architectures proposed the use of Siamese or dual-stream networks as the backbone \cite{qin2018joint,qiustacom19}. These models utilized separate encoder branches to extract features from the moving and fixed images, which were then fused and passed through the decoder to generate the deformation. Similar to U-Net style networks, Siamese and dual networks also used a contracting path for image encoding and an expansive path for decoding the deformations from the encoded image features.

There are several other extensions which have been shown to improve registration performance, including the use of different loss functions and regularization techniques, the incorporation of prior knowledge, and the estimation of deformation using B-spline control points. For instance, the segmentation Dice loss \cite{balakrishnan2019voxelmorph,ha2020semantically} measures the overlap between segmented regions of fixed and moving images and has been used to improve registration performance between different anatomical structures. Regularization techniques, such as inverse consistency \cite{zhang2018inverse} and squaring and scaling \cite{dalca2018unsupervised}, are utilized to encourage diffeomorphic deformations. These methods are particularly useful for task-specific registration. In addition, combining deep learning with conventional model-based approaches \cite{jia2021learning, qiu2022embedding} and incorporating prior knowledge \cite{fan2019birnet} have also demonstrated improvements to registration performance. For example, Jia et al.\cite{jia2021learning} used the variational model to linearize the bright consistency similarity metric and optimized the auxiliary variable with a cascaded U-Net. Albeit slightly slower, their approach was found to be more accurate than both single (VoxelMorph \cite{balakrishnan2019voxelmorph}) and cascaded U-Net (RC-Net \cite{Zhao_2019_ICCV}) for cardiac motion estimation. Another line of work is to estimate a grid of B-spline control points with regular spacing \cite{de2019deep,qiu2021learning}, which is then interpolated based on cubic B-spline basis functions \cite{rueckert1999nonrigid, rueckert2006diffeomorphic, duan2019automatic}. These networks predict deformations more efficiently by estimating only a few control points, although currently they are less accurate.

\begin{figure*}[t]
     \centering
     \includegraphics[width=0.98\textwidth]{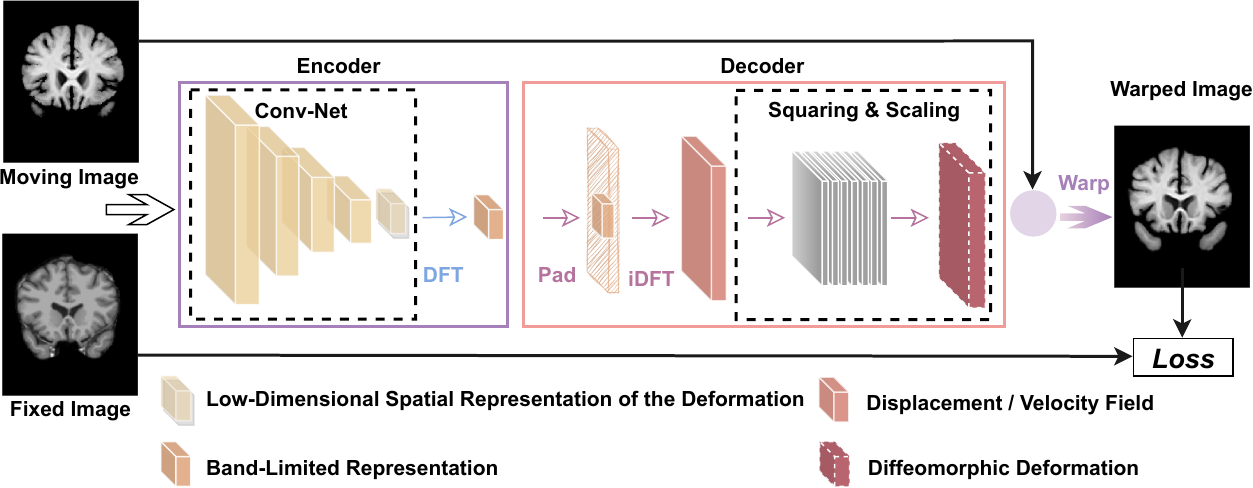}
     \caption{Architecture of our unsupervised Fourier-Net. It contains 1) a convolutional encoder that first produces a low-dimensional representation of a displacement or velocity field, followed by an embedded discrete Fourier transformation (DFT) layer to map this low-dimensional representation into the band-limited Fourier domain; 2) a parameter-free model-driven decoder that adopts a zero-padding layer, an inverse DFT (iDFT) layer, and 7 optional squaring and scaling layers to reconstruct the displacement field or deformation into the full-resolution spatial domain from its band-limited Fourier domain; 3) a warping layer to deform the moving image; and 4) a learning objective loss function.  }
     \label{fig:flowchart1}
 \end{figure*}

\textbf{Supervised methods:} Instead of minimizing a similarity metric and regularization term as per unsupervised methods, supervised approaches learn from ground truth deformation fields. However, they have several pitfalls: 1) it is generally hard to provide human-annotated ground truth deformations for supervision; and 2) if trained using numerical solutions of other iterative methods, the performance of these supervised registration methods may be limited by iterative methods. Yang et al. proposed Quicksilver \cite{yang2017quicksilver} which is a supervised encoder-decoder network and trained using the initial momentum of LDDMM as the supervision signal. Wang et al. extended Flash \cite{zhang2019fast} to DeepFlash \cite{wang2020deepflash} in a learning framework in lieu of iterative optimization. Compared to Flash, DeepFlash accelerated the computation of the initial velocity fields but still needed to solve a PDE (i.e., EPDiff equation) in the Fourier domain so as to recover the full-resolution deformation in the spatial domain, which was shown to be slow. The fact that DeepFlash required the numerical solutions of Flash \cite{zhang2019fast} as training data attributes to lower registration performance than Flash.

Although DeepFlash also learns a low-dimensional band-limited representation, it differs from our Fourier-Net in four aspects, which represent our novel contributions to this area. First, DeepFlash is a supervised method that requires ground truth velocity fields calculated from Flash (30 minutes per 3D image pair in CPU) prior to training, whilst Fourier-Net is a simple and effective unsupervised method thanks to our proposed model-driven decoder. Second, DeepFlash is a multi-step method whose network's output requires an additional PDE algorithm \cite{zhang2019fast} to compute final full-resolution spatial deformations, whilst Fourier-Net is a holistic model that can be trained and used in an end-to-end manner. Third, DeepFlash requires two individual convolutional networks to estimate real and imaginary signals in the band-limited Fourier domain, whilst Fourier-Net uses only one single network directly mapping image pairs to a reduced-resolution displacement field without the need of complex-valued operations. Lastly, DeepFlash is essentially an extension of Flash and it is difficult for the method to benefit from vast amounts of data, whilst Fourier-Net is flexible and can easily learn from large-scale datasets.

\section{Proposed Methods}
In this section, we first introduce Fourier-Net, and its extended versions: Fourier-Net+ and cascaded Fourier-Net+. We then detail their network architectures and finally present their loss functions used in our experiments. 
\subsection{Fourier-Net}
\label{sec:fnet}
We illustrate Fourier-Net in Figure~\ref{fig:flowchart1}, where its encoder takes a pair of spatial images as input and encodes them to a low-dimensional representation of the displacement field (or velocity field if diffeomorphisms are imposed) in the band-limited Fourier domain. The decoder then brings the displacement field (or velocity field) from the band-limited Fourier domain to the spatial domain via a padding and an iDFT. The decoder ensures that the input and output have the same spatial size. Next, squaring and scaling layers are optionally used to encourage a diffeomorphism in the final deformation. Finally, by minimizing the loss function, the warping layer deforms the moving image to be similar to the fixed image.

\textbf{Encoder of Fourier-Net}: the encoder aims to learn a displacement or velocity field in the band-limited Fourier domain, which requires the encoder to handle complex-valued numbers. One may directly use complex convolutional networks \cite{trabelsi2017deep}, which were designed for the case where both input and output are complex values. Note that complex convolutional operations sacrifice computational efficiency. Instead, DeepFlash \cite{wang2020deepflash} tackles this problem by first converting input image pairs to the Fourier domain and then using two individual real-valued convolutional networks to learn the real and imaginary signals separately. However, such an approach makes training and inference costly.

\begin{figure}[t]
    \centering
    \includegraphics[width=0.46\textwidth]{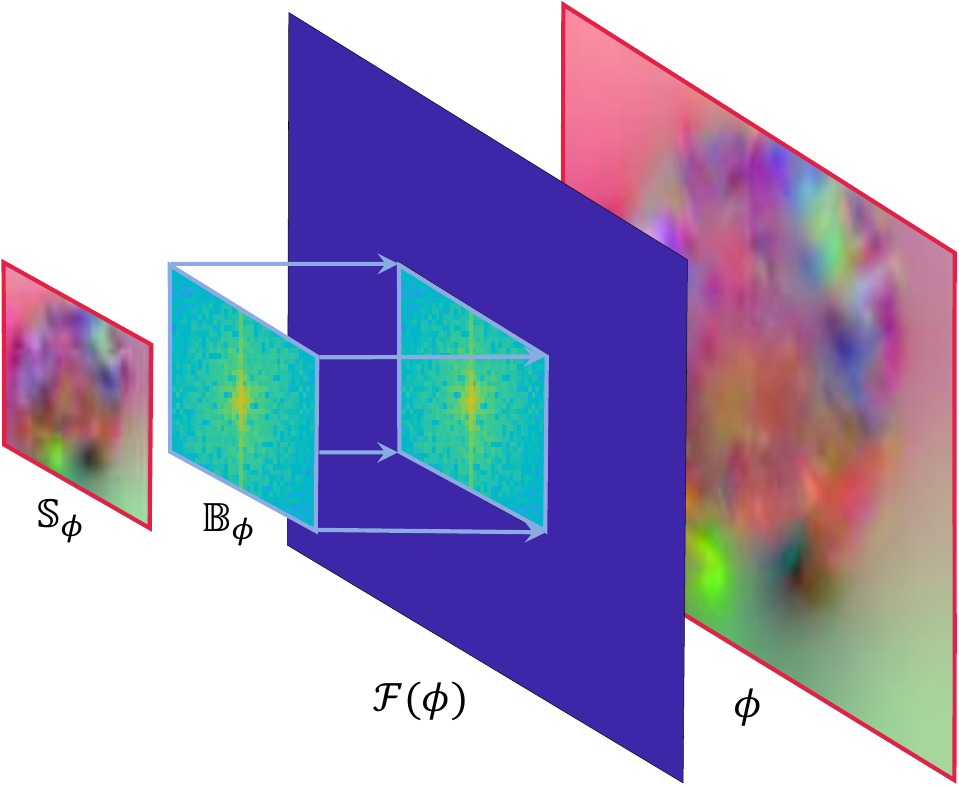}
    \caption{Connection between low-dimensional spatial displacement field $\mathbb{S}_{\boldsymbol{\phi}}$, band-limited Fourier coefficients $\mathbb{B}_{\boldsymbol{\phi}}$, full-resolution Fourier coefficients $\mathcal{F}({\boldsymbol{\phi}})$ by zero-padding $\mathbb{B}_{\boldsymbol{\phi}}$, and full-resolution displacement field $\boldsymbol{\phi}$ by taking iDFT of $\mathcal{F}(\boldsymbol{\phi})$.} \label{fig:large_small_spatial}
\end{figure}
 
To bridge the domain gap between real-valued spatial images and complex-valued band-limited displacement fields without increasing complexity, we propose to embed a DFT layer at the end of the encoder. This is a simple and effective way to produce complex-valued band-limited displacement fields without the network handling complex values itself. Let us denote the moving image as $I_M$, the fixed image as $I_F$, the convolutional network as $\boldsymbol{CNN}$ with the parameters ${\Theta^1}$, the DFT layer as $\mathcal{F}$, the full-resolution spatial displacement field as $\boldsymbol{\phi}$, and the complex band-limited displacement field as $\mathbb{B}_{\boldsymbol{\phi}}$. We therefore define our encoder as $\mathbb{B}_{\boldsymbol{\phi}} = {\mathcal F}({\boldsymbol{CNN}}(I_M, I_F; {{\Theta^1}}))$, resulting in a compact, efficient implementation in contrast to DeepFlash \cite{wang2020deepflash}.  

We also attempted to regress $\mathbb{B}_{\boldsymbol{\phi}}$ directly from $I_M$ and $I_F$ using only convolutional layers, i.e., $\mathbb{B}_{\boldsymbol{\phi}} = {\boldsymbol{CNN}}(I_M, I_F; {{\Theta^1}})$. However, our experimental results suggested that this is very difficult for the network to learn, resulting in lower performance as detailed in Table \ref{tab:ablation_2doasis_dft}. If the CNN is required to directly learn a band-limited displacement field, it must go through two domains in total: first mapping the spatial images to the spatial displacement field and then mapping this displacement field into its band-limited Fourier domain. In this case, we believe the domain gap is too big for a CNN to learn such a mapping. Our network however only needs to go through one domain and then the DFT layer ($\mathcal{F}$) handles the second domain. Experimentally, we found this approach to be more effective.

So far, we have given an intuitive explanation of how the encoder in our network learns. We now discuss the mathematical relationship between the low-dimensional spatial displacement field $\mathbb{S}_{\boldsymbol{\phi}}= {\boldsymbol{CNN}} (I_0, I_1; {{\Theta^1}})$, its band-limited representation $\mathbb{B}_{\boldsymbol{\phi}}=\mathcal{F}(\mathbb{S}_{\boldsymbol{\phi}})$, as well as the displacement field $\boldsymbol{\phi}$ in the full-resolution spatial domain (see Figure~\ref{fig:large_small_spatial} for details). For simplicity, we use a 2D displacement field as an example and the formulations below can be easily extended to 3D cases. A define a general DFT used on $\boldsymbol{\phi}$ as follows:
\begin{equation} \label{eq:dft}
[{\mathcal {F}}(\boldsymbol{\phi})]_{k, l}= \sum_{i=0}^{H-1} \sum_{j=0}^{W-1} \boldsymbol{\phi}_{i, j} e^{-\sqrt{-1} \left(\frac{2 \pi k}{H} i+\frac{2 \pi l}{W} j\right)},
\end{equation}
where $\boldsymbol{\phi}$ is of size $H\times W$, $i \in [0, H-1]$ and $j\in [0, W-1]$  are the discrete indices in the spatial domain, and $k \in [0, H-1]$ and $l \in [0, W-1]$ are the discrete indices in the frequency domain.
In our Fourier-Net, $\boldsymbol \phi$ is actually a low-pass filtered displacement field. If we define a $H \times W$ sized sampling mask $\mathcal D$ whose entries are zeros if they are on the positions of high-frequency signals in $\boldsymbol \phi$ and ones if they are on the low-frequency positions. With $\mathcal D$, we recover the displacement field $\boldsymbol{\phi}$ from Eq.~\eqref{eq:dft} as follows:
\begin{equation} \label{eq:idft}
 {\boldsymbol{\phi}}_{i, j} = {\frac{1}{HW}}\sum_{k=0}^{H-1} \sum_{l=0}^{W-1} {\mathcal{D}}_{k,l} [{\mathcal {F}}(\boldsymbol{\phi})]_{k, l} e^{\sqrt{-1} \left(\frac{2 \pi i}{H} k+\frac{2 \pi j}{W} l\right)}.
\end{equation}
If we shift all low-frequency signals of $\boldsymbol{\phi}$ to a center patch of size $\frac{H}{a}\times\frac{W}{b}$ ($\frac{H}{a},\frac{W}{b},a=2Z_a,b=2Z_b,Z_a,Z_b \in \mathbb{Z}^+$), center-crop the patch (denoted by $\mathbb{B}_{\boldsymbol{\phi}}$), and then perform the iDFT on this patch, we obtain $\mathbb{S}_{\boldsymbol{\phi}}$ in Eq.~\eqref{eq:dft2}:
\begin{equation} \label{eq:dft2}
[\mathbb{S}_{\boldsymbol{\phi}}]_{\widehat{i}, \widehat{j}} = \frac{ab}{H W}  \sum_{\widehat{k}=0}^{\frac{H}{a}-1} \sum_{\widehat{l}=0}^{\frac{W}{b}-1} [\mathbb{B}_{\boldsymbol{\phi}}]_{\widehat{k},\widehat{l}} e^{\sqrt{-1} \left(\frac{2 \pi a\widehat{i}}{H} \widehat{k}+\frac{2 \pi b\widehat{j}}{W} \widehat{l}\right)},
\end{equation}
where $\widehat{i} \in [0, \frac{H}{a}-1]$ and $\widehat{j}\in [0, \frac{W}{b}-1]$ are the indices in the spatial domain, and $\widehat{k} \in [0, \frac{H}{a}-1]$ and $\widehat{l} \in [0, \frac{W}{b}-1]$ are the indices in the frequency domain. Note that $\mathbb{S}_{\boldsymbol{\phi}}$ is a low-dimensional spatial representation of $\boldsymbol \phi$ and we are interested in their mathematical connection. Another note is that $\mathbb{S}_{\boldsymbol{\phi}}$ actually contains all the information of its band-limited Fourier coefficients in $\mathbb{B}_{\boldsymbol{\phi}}$. As such, we do not need the network to learn the coefficients in $\mathbb{B}_{\boldsymbol{\phi}}$ and instead only to learn its real-valued coefficients in $\mathbb{S}_{\boldsymbol{\phi}}$.

Since most of entries ($\frac{a \times b -1}{a \times b}\%$) in $\mathcal {F}(\boldsymbol{\phi})$ are zeros, and the values of remaining entries are exactly the same as in $\mathbb{B}_{\boldsymbol{\phi}}$, we can conclude that $\mathbb{S}_{\boldsymbol{\phi}}$ contains all the information $\boldsymbol{\phi}$ can provide, and their mathematical connection is
\begin{equation}
[\mathbb{S}_{\boldsymbol{\phi}}]_{\widehat{i}, \widehat{j}}  =  ab \times \boldsymbol{\phi}_{a \widehat{i}, b\widehat{j}}.
\end{equation}
With this derivation, we show that we can actually recover a low-dimensional spatial representation $\mathbb{S}_{\boldsymbol{\phi}}$ from its full-resolution spatial displacement field $\boldsymbol{\phi}$, as long as they have the same low-frequency coefficients $\mathbb{B}_{\boldsymbol{\phi}}$. This shows that there exists a unique mapping function between $\mathbb{S}_{\boldsymbol{\phi}}$ and $\boldsymbol{\phi}$ and that it is reasonable to use a network to learn $\mathbb{S}_{\boldsymbol{\phi}}$ directly from image pairs.

\textbf{Model-driven decoder:} The decoder contains no learnable parameters. We instead replace the expansive path with a zero-padding layer, an iDFT layer, and an optional squaring and scaling layer. The output from the encoder is a band-limited representation $\mathbb{B}_{\boldsymbol{\phi}}$. To recover the full-resolution displacement field $\boldsymbol{\phi}$ in the spatial domain, we first pad the patch $\mathbb{B}_{\boldsymbol{\phi}}$ containing mostly low-frequency signals to the original image resolution with zeros (i.e., ${\mathcal {F}}(\boldsymbol{\phi})$). We then feed the zero-padded complex-valued coefficients ${\mathcal {F}}(\boldsymbol{\phi})$ to an iDFT layer consisting of two steps: shifting the Fourier coefficients from centers to corners and then applying the standard iDFT to convert them into the spatial domain. The output from Fourier-Net is thus a full-resolution spatial displacement field. An illustration of this process is given in Figure~\ref{fig:large_small_spatial}. Both padding and iDFT layers are differentiable and therefore Fourier-Net can be optimized via standard back-propagation.



We also propose a diffeomorphic variant of Fourier-Net which we term Diff-Fourier-Net. A diffeomorphic deformation is defined as a smooth and invertible deformation, and in Diff-Fourier-Net we need extra squaring and squaring layers for the purpose. The output of the iDFT layer can be regarded as a stationary velocity field denoted by $\boldsymbol{v}$ instead of the displacement field $\boldsymbol \phi$. In group theory, $\boldsymbol{v}$ is a member of Lie algebra, and we can exponentiate this stationary velocity field (i.e., $Exp(\boldsymbol{v}$)) to obtain a diffeomorphic deformation. In this paper, we use seven scaling and squaring layers \cite{ashburner2007fast,dalca2018unsupervised} to impose such a diffeomorphism.

\subsection{Fourier-Net+}
\label{sec:fnet+}
In Fourier-Net, we proposed a model-driven decoder to replace some of the expansive convolutional layers in a U-Net style backbone. By doing so, Fourier-Net removed the need of progressively decoding the displacement/velocity field from the latent features learned from the encoder. Such a decoder is thus capable of reducing the computational cost and improving inference speed. However, Fourier-Net still takes the original full-resolution image pairs as input. For 3D images, the encoder of Fourier-Net involves multiple 3D convolutional layers, which are computationally expensive operations, particularly in the earlier layers of the model. To further reduce the computational cost and memory footprint, we propose Fourier-Net+ by embedding a model-driven encoding layer before the contracting convolutional layers.

\begin{figure*}[t]
     \centering
     \includegraphics[width=0.98\textwidth]{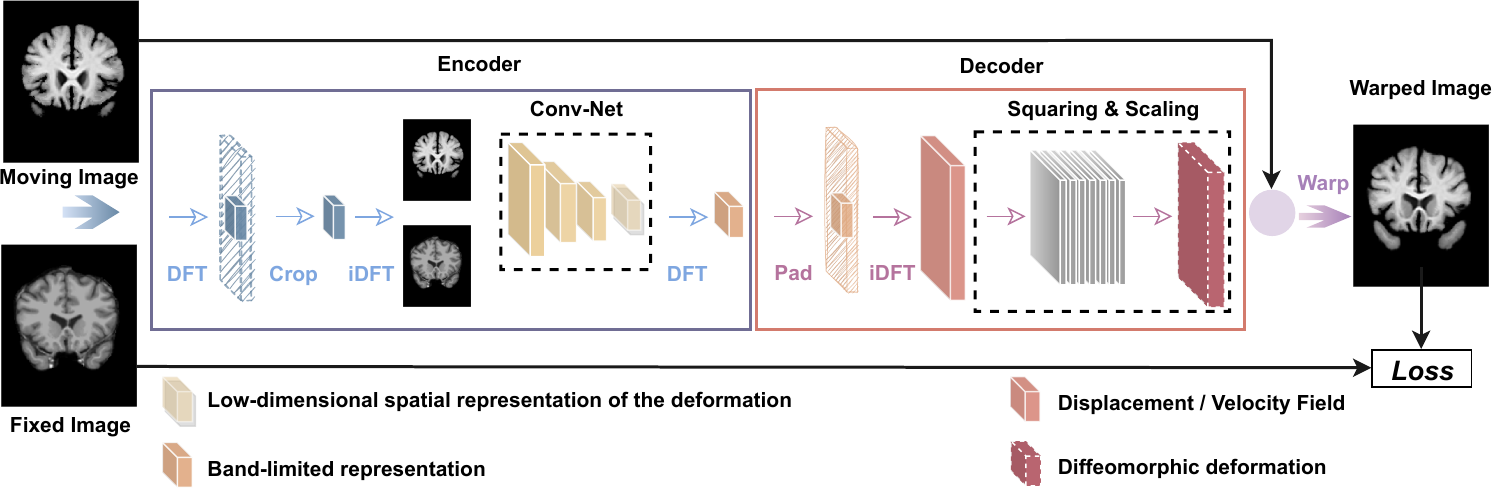}
     \caption{Architecture of our Fourier-Net+. The early layers from the expansive path in Fourier-Net are replaced with the embedded DFT, Crop, and iDFT layers. This model-driven encoder directly reduces the resolution of the input images, which are then fed through to the Conv-Net.}
     \label{fig:flowchart2}
 \end{figure*}
 
\begin{figure}[t]
    \centering \includegraphics[width=0.46\textwidth]{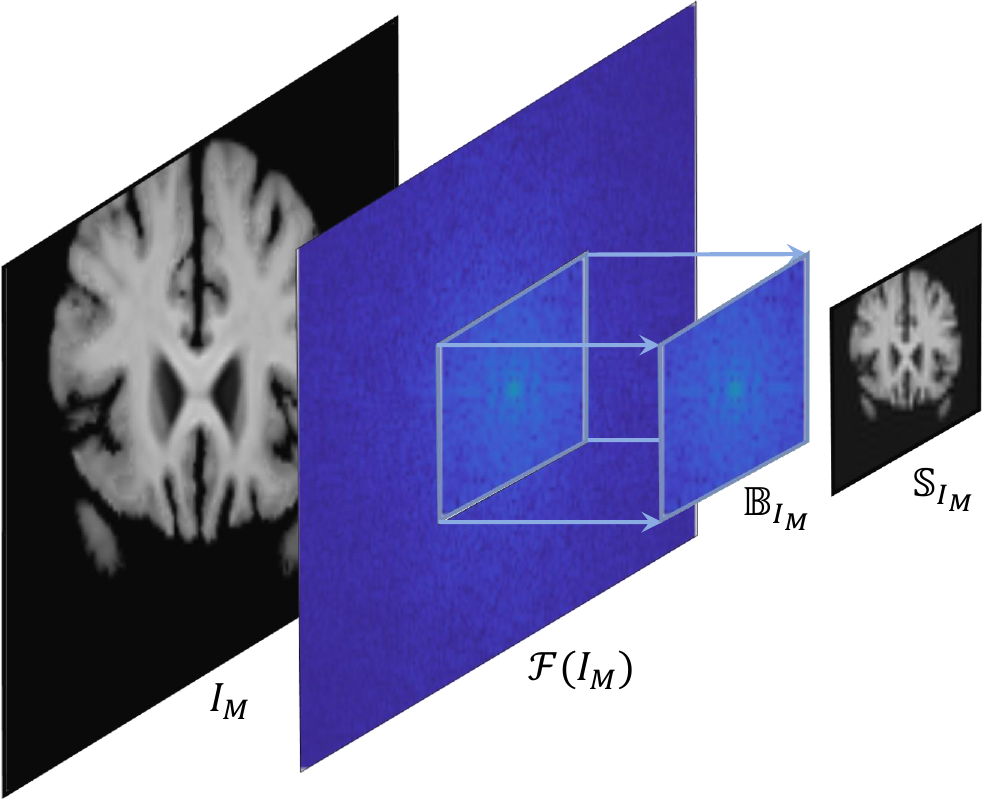}
    \caption{Instead of using the original full-resolution images $I_M$ and $I_F$ as input to our network, we use the band-limited representation of the images $\mathbb{S}_M$ and $\mathbb{S}_F$. For example, in case of $I_M$ we first transform the image into the Fourier domain $\mathcal{F}({I_M})$ with a DFT layer, followed by applying a center-cropping to produce the low-frequency signals $\mathbb{B}_M$. Since the low-frequency $\mathbb{B}_M$ is in the complex-valued Fourier space, we use an inverse DFT operation to convert the low-frequency patch $\mathbb{B}_M$  to the spatial domain $\mathbb{S}_M$. The spatial patch $\mathbb{S}_M$ is the band-limited representation of the original image, which is then fed into conv-net.} \label{fig:large_small_spatial_f+}
\end{figure}


\textbf{Model-driven encoder:} It is more efficient for learning if we feed our encoder a band-limited representation of input images, which we term band-limited images in line with band-limited displacements. With this consequently much lighter convolutional network, we are able to estimate their band-limited displacements from these band-limited images and further reduce computional costs. Similarly to our approach with the decoder in Fourier-Net, in Fourier-Net+ the input images $I_M$ and $I_F$ are first mapped into the frequency domain using a DFT layer, forming $\mathcal{F}{(I_M)}$ and $\mathcal{F}{(I_F)}$. We then perform center-cropping on the low-frequency regions, forming $\mathbb{B}_{I_M}$ and $\mathbb{B}_{I_F}$, which are transformed to the spatial domain using iDFT. The real numbers are then taken, and the resulting spatial patches $\mathbb{S}_{I_M}$ and $\mathbb{S}_{I_F}$ are the band-limited images, as illustrated in Figures~\ref{fig:large_small_spatial_f+}. Once we have the band-limited images, we only need a small $\boldsymbol{CNN}$ parameterized by ${{\Theta}^2}$ to estimate their band-limited displacements, i.e., $\mathbb{S}_{\phi} = \boldsymbol{CNN}(\mathbb{S}_{I_M}, \mathbb{S}_{I_F}; {{\Theta}^2})$. This encoder removes several convolutional layers in the contracting path of the Fourier-Net encoder, which we expect to further accelerate the registration process and reduce the memory footprint of Fourier-Net significantly.  


The architecture of Fourier-Net+ is shown in Figure~\ref{fig:flowchart2}. Except for the incorporation of a model-driven encoding layer, the remaining parts of Fourier-Net+ are exactly the same as Fourier-Net. By adding the squaring and scaling layers at the end of Fourier-Net+, we get its diffeomorphic version, which we term Diff-Fourier-Net+. An important note here is that the warping layer in Fourier-Net+ or Diff-Fourier-Net+ warps the originally sized images instead of the band-limited images in order to compute the final loss.


\subsection{Cascaded Fourier-Net+}
\label{sec:kfnet}

\begin{figure*}[t]
     \centering     \includegraphics[width=0.98\textwidth]{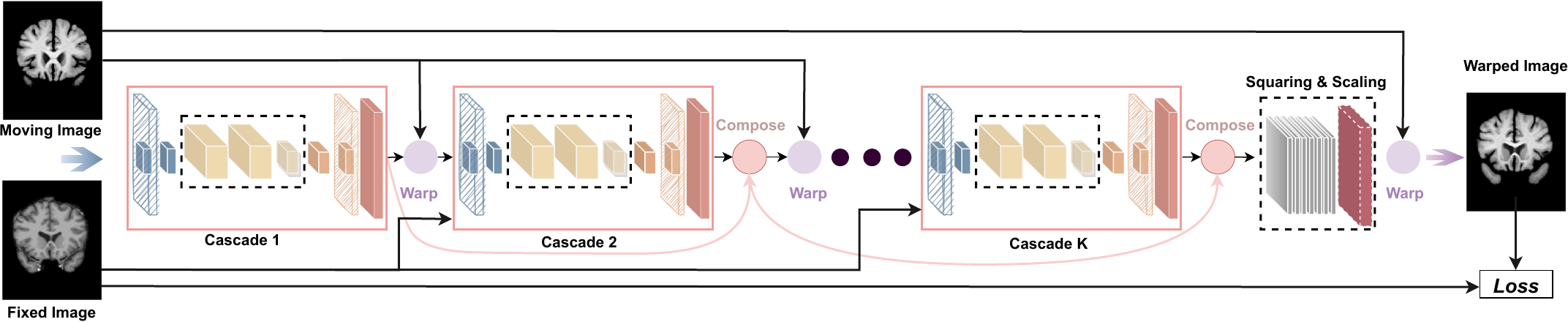}
     \caption{Architecture of the proposed Cascaded Fourier-Net+ with $K$ cascades (denoted $K\times$Fourier-Net+ in results tables). Each cascade contains a single Fourier-Net+, where the final deformation is composed of the deformations from each individual cascade. The network is trained end-to-end using only a single loss function following the final cascade or squaring and scaling layer if present.}
     \label{fig:flowchart3}
 \end{figure*}
 
Due to the band-limited representation of both images and deformations, Fourier-Net+ is lighter than Fourier-Net in terms of the number of parameters and computations. However, such a light network may face limitations in accurately capturing complex deformations (e.g., in brain images). To tackle this problem, we propose a cascaded version of Fourier-Net+, which is illustrated in Figure~\ref{fig:flowchart3}. For the diffeomorphic version of this network, we impose the squaring and scaling layers after the last cascade. The loss function is also used after the last cascade. For terminology, $K\times$Fourier-Net+ means that we use a Fourier-Net+ with $K$ cascades, and Diff-$K\times$Fourier-Net+ is the diffeomorphic version of $K\times$Fourier-Net+. This cascaded Fourier-Net+ is parameterized by $\Theta^3$, and its network parameters are not shared across different cascades.


Technically, in the first cascade of $K\times$Fourier-Net+ the input is the moving image $I_M$ and the fixed image $I_F$. From the second cascade onward, the input is $I_{M}^{w(k)}$ (a warped version of $I_M$, $k \in [1,K]$) and $I_F$. The process iterates until Cascade $k$, where the input is $I_{M}^{w(k-1)}$ and $I_F$ and the output $\delta \phi^{(k)}$. Specifically, as in \cite{Zhao_2019_ICCV}, $I_M^{w(k)}$ is defined as:
\begin{equation}
    I_M^{w(k)} = ((((I_M \circ \delta\phi^{(1)})\circ \delta\phi^{(2)}) \circ \dotsb) \circ \delta\phi^{(k-1)}) \circ \delta\phi^{(k)}, 
\end{equation}
or equivalently
\begin{equation}
    I_M^{w(k)} = I_M\circ \boldsymbol{\phi}^{(k)},
\end{equation}
where $\boldsymbol{\phi}^{(k)}$ is the displacement field computed by composing the outputs from Cascade 1 to $k$:
\begin{equation}
    \boldsymbol{\phi}^{(k)} = \delta\phi^{(1)} \circ \delta\phi^{(2)} \circ \dotsb \circ \delta\phi^{(k-1)} \circ \delta\phi^{(k)}.
\end{equation}
In the case of $k=K$, $\boldsymbol{\phi}^{(K)}$ denotes the final output displacement field used to warp the original moving image $I_M$ in the computation of the loss.

\subsection{Network Architectures and Loss Functions}
\begin{figure*}[t]
    \centering
    \includegraphics[width=0.99\textwidth]{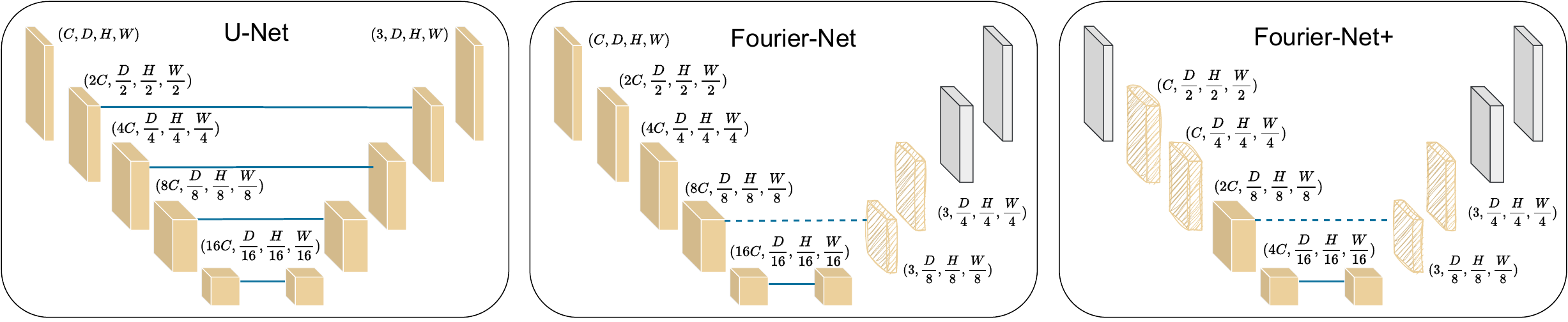}
    \caption{
    The CNN architectures used for U-Net, Fourier-Net, and Fourier-Net+ in our experiments. Here, $(C, D, H, W)$ represents the channels, depth, height, and width of a layer. Fourier-Net is based on a U-Net, where grey layers in the decoder are removed. The shaded layers are optionally used to control the resolution of the band-limited deformation. Fourier-Net+ additionally forgoes the first layer of the U-Net, with the following two optional layers corresponding to the choice of band-limited image resolution. Optimal choices for band-limited images and displacements were determined through our experimentation.}
  \label{fig:convnet}
\end{figure*}

\textbf{Network architectures:}  For all networks using a U-Net backbone in our experiments, we employed the architecture defined in Figure~\ref{fig:convnet}, which is similar to that used in \cite{zhang2018inverse,Mok_2020_CVPR,jia2022u}. In this network, there are five convolutional layers in the contracting path and five convolutional layers in the expansive path. Specifically, given a pair of moving and fixed images in 3D, each with a size of $D\times H \times W$, the layer size in the network flows as $(C,D,H,W) \rightarrow(2C,\frac{D}{2},\frac{H}{2},\frac{W}{2}) \rightarrow(4C,\frac{D}{4},\frac{H}{4},\frac{W}{4})\rightarrow(8C,\frac{D}{8},\frac{H}{8},\frac{W}{8})\rightarrow(16C,\frac{D}{16},\frac{H}{16},\frac{W}{16})$ in the contracting path. These layers in the expansive path were progressively upsampled to $(3,D,H,W)$, which is the size of the final displacement/velocity field. The architecture of Fourier-Net given in Figure~\ref{fig:convnet} was modified from the U-Net backbone by discarding several layers in the expansive path. In the two variants of Fourier-Net, we  used fewer layers in the expansive path, which lead to a smaller spatial size of the output (last) layer and reduces the convolutional computations in higher-dimensional space. A smaller size of the output layer rapidly decreases the model parameters and speeds up training and inference time, but may lead to lower performance. A larger size of output layer will retain registration accuracy but eliminate the efficiency advantage of our methods. We noticed that different datasets favor different sizes and investigated in our experiments two different sizes for the output layer, i.e., $(3,\frac{D}{8},\frac{H}{8},\frac{W}{8})$ and $(3,\frac{D}{4},\frac{H}{4},\frac{W}{4})$. As illustrated by Figure~\ref{fig:convnet}, in Fourier-Net+ we instead used small-sized band-limited images as input, allowing us to remove some convolutional layers in the contracting path to further save the computational cost. In our experiments we studied two different sizes of such band-limited images, i.e., $(\frac{D}{2},\frac{H}{2},\frac{W}{2})$ and $(\frac{D}{4},\frac{H}{4},\frac{W}{4})$, which in combination with the two sizes of band-limited displacements results in a total of four Fourier-Net+ variants. 


\textbf{Loss functions:} For Fourier-Net, Fourier-Net+, and $K\times$Fourier-Net+, the final output is the full-resolution displacement field $\boldsymbol{\phi}$ (in $K\times$Fourier-Net+ we let $\boldsymbol{\phi}=\boldsymbol{\phi}^K$). Both warping layers in 2D and 3D are based on linear interpolation as in \cite{jaderberg2015spatial,balakrishnan2019voxelmorph}. We define an unsupervised training loss ${\mathcal L}({\boldsymbol{\Theta}})$, computed from the moving image $I_M$, the fixed image $I_F$, the predicted displacement field $\boldsymbol{\phi}$, and the network parameters ${\boldsymbol{\Theta}}$. In detail, ${\mathcal L}({\boldsymbol{\Theta}})$ is of the following form:
\begin{equation}      \frac{1}{N} \sum_{i=1}^N {\mathcal L}_{S}( I_{M_i} \circ  ({\boldsymbol{\phi}_i}({\boldsymbol{\Theta}}) + {\rm{Id}}) - I_{F_i} )  + \frac{\lambda}{N} \sum_{i=1}^N \|\nabla \boldsymbol{\phi}_i({\boldsymbol{\Theta}}) \|_2^2 ,
\end{equation}
where $N$ is the number of training pairs, ${\rm{Id}}$ is the identity grid, $\circ$ is the warping operator, and $\nabla$ is the first order gradient implemented using finite differences \cite{lu2016implementation,duan2016edge}. The first term ${\mathcal L}_{S}$ defines the similarity between warped moving images and fixed images, and the second term defines the smoothness of displacement fields. Here $\lambda$ is a hyper-parameter balancing the two terms. 

For all diffeomorphic variants of Fourier-Net and Fourier-Net+, the final output is the exponentiated full-resolution velocity field $\boldsymbol{v}$ after the squaring and squaring layers. The training loss ${\mathcal L}({\boldsymbol{\Theta}})$ in this case is defined as
\begin{equation}
     \frac{1}{N} \sum_{i=1}^N  {\mathcal L}_{S}( I_{M_i} \circ  Exp(\boldsymbol{v}_i({\boldsymbol{\Theta}})) - I_{F_i} )  + \frac{\lambda}{N} \sum_{i=1}^N \|\nabla \boldsymbol{v}_i({\boldsymbol{\Theta}}) \|_2^2.
\end{equation}
${\mathcal L}_{S}$ can be either mean squared error (MSE) or normalized cross-correlation (NCC), which we clarify in our experiments. Depending the network architecture, ${\boldsymbol{\Theta}}$ is either $\Theta^1$, $\Theta^2$, or $\Theta^3$ (see Sec.~\ref{sec:fnet}, \ref{sec:fnet+}, and \ref{sec:kfnet} for details). The aim is to minimize ${\mathcal L}({\boldsymbol{\Theta}})$ with respect to ${\boldsymbol{\Theta}}$ using gradient descent via backprogration. 

\section{Experiments}
In this section, we detail the datasets used in our experimentation, provide implementation details and conduct ablation and parameter studies to demonstrate the utility of our contributions. We finally compare our method with a range of state-of-the-art methods across three different registration tasks.

\subsection{Datasets}

\textbf{OASIS-1 dataset} \cite{marcus2007open} consists of a cross-sectional collection of T1-weighted brain MRI scans from 416 subjects. In experiments, we used the pre-processed OASIS data provided by the Learn2Reg challenge\cite{hoopes2021hypermorph} and performed subject-to-subject brain registration. This dataset has 414 2D $160\times192$ slices and masks extracted from their corresponding 3D $160\times192\times224$ volumes. We randomly split this 2D dataset into 201, 12, and 201 images for training, validation, and testing. After pairing, we used 40200, 22, and 400 image pairs for training, validation, and testing, respectively. Each 2D segmentation mask  contained 24 automated labels from FreeSurfer. Further details of this pre-processed data can be found at the MICCAI 2021 Learn2Reg challenge\footnote{\url{https://learn2reg.grand-challenge.org/Learn2Reg2021/}}. 

 
\textbf{IXI dataset}\footnote{\url{https://brain-development.org/ixi-dataset/}} contains nearly 600 MRI scans from healthy subjects. In experiments, we used the pre-processed IXI data provided by \cite{chen2021transmorph} to perform atlas-based brain registration. The atlas is generated by the authors of  \cite{chen2021transmorph} using the method in \cite{kim2021cyclemorph}. There are in total 576 $160\times192\times224$ volumetric images in this pre-processed dataset, which are split into 403 for training, 58 for validation, and 115 for testing. There is no pairing step for this dataset as it is an atlas-to-subject registration task. The performance of this task was evaluated with 30 labeled anatomical structures\footnote{\url{https://github.com/junyuchen245/TransMorph_Transformer_for_Medical_Image_Registration/blob/main/IXI/TransMorph_on_IXI.md}}.


{\textbf{3D-CMR dataset}} \cite{duan2019automatic,thorley2021nesterov,jia2021learning} consists of 220 pairs of 3D high-resolution cardiac MRI scans, in which each scan is captured during only one single breath-hold and includes the End-diastolic (ED) to End-systolic (ES) frames of the cardiac cycle. In our experiments, we re-sampled all scans from original resolution  1.2$\times$1.2$\times$2.0$mm^3$ to 1.2$\times$1.2$\times$1.2$mm^3$ and center-cropped 128$\times$128$\times$96 sized volumes. We randomly split the data into 100, 20, and 100 corresponding to training, validation, and testing sets. The Dice score and Hausdorff distance (HD) between the warped ES segmentation and the ED ground truth mask was measured on three anatomical structures: left ventricle cavity (LV), left ventricle myocardium (Myo), and right ventricle cavity (RV). We performed motion estimation from an ES state to an ED state from a cardiac cycle.

\subsection{Implementation Details} 
The U-Net backbone used in experiments is given in Figure~\ref{fig:convnet}. There were 5 blocks in both the contracting and expansive paths. In the encoder, the first block directly encoded the input image pair to $C$ feature maps, each with a size of $D\times H \times W$. For the remaining 4 blocks, each contained 2 sequential convolutional layers, where the first layer maintained the same spatial resolution as its input, and the second layer performed a down-sampling with a stride of 2 and doubled the number of feature channels. In the decoder, each of the first 4 blocks contained a fractionally-strided convolutional layer followed by 2 sequential convolutional layers, where the fractionally-strided convolutional layer performed an up-sampling with a stride of 2, and the convolutional layers halved the number of feature channels.  The output from the last block was the final displacement field. The kernel size in all blocks was $3\times 3\times 3$, and each convolution was followed by a PReLU activation except the last output layer, which did not have any activation function.

We implemented all our proposed networks (see Figure~\ref{fig:convnet} middle and right) using PyTorch, where training is optimized using Adam. To adapt to 2D images, 3D kernels were changed to 2D, each with a size of $3 \times 3$. For training in both 2D and 3D, we tuned built-in hyper-parameters on a held-out validation set. In terms of loss functions, we used MSE to train our networks on OASIS-1 for 10 epochs, and the optimal performance was achieved with $\lambda=0.01$ for all our networks. On IXI, we trained Fourier-Net, Diff-Fourier-Net, Fourier-Net+ 
 and $K\times$Fourier-Net+ with the NCC loss for 1000 epochs with $\lambda=5$. In contrast, $K\times$Fourier-Net+ and Diff-$K\times$Fourier-Net+ are trained optimally with $\lambda=2$. On 3D-CMR, we used MSE to train our networks for 1000 epochs with $\lambda=0.001$. All our networks were trained using an NVIDIA A100 GPU.

\subsection{Ablation Studies} 
\label{subsec:ablation_2doasis}
In this section, we detail our ablation studies where we investigated whether the proposed modules in Fourier-Net and its variants were effective. All experiments undertaken in this section were conducted on OASIS-1.

\begin{table}[]
\centering
\caption{Comparisons between  Fourier-Net and Fourier-Net without an embedded DFT in the encoder. 'Output' denotes the output resolution of the $\boldsymbol{CNN}$ encoder.  DFT indicates embedding the DFT layer in the encoder. We present the \textit{Dice} coefficient, percentage of negative Jacobian determinant of deformation ($|J|_{< 0}\%$ ), the number of \textit{Mult-Adds} operations in millions required for inference, and the \textit{Memory} footprint in Megabyte (MB) required for one forward/backward for each permutation of DFT and encoder output resolution.}
\label{tab:ablation_2doasis_dft}
\begin{tabular}{cccccc}
\hline
DFT & Output       & Dice$\uparrow$ & $|J|_{< 0}\%$       & Mult-Adds & Memory \\ \hline
\xmark     & $20\times 24$ & 0.664$\pm$0.040 & 0.159±0.207  & 890.55        & 33.53       \\
\cmark     & $20\times 24$ & 0.732$\pm$0.042 & 0.434±0.355 & 679.19        & 31.18       \\\hline
\xmark     & $40\times 48$ & 0.675$\pm$0.038 & 0.279±0.257 & 1310.0          & 42.95       \\
\cmark     & $40\times 48$ & 0.756$\pm$0.039 & 0.753±0.408 & 888.25        & 35.89      \\\hline
\end{tabular}
\end{table}

\textbf{Impact of embedding a DFT layer:} In Table \ref{tab:ablation_2doasis_dft}, we show the necessity of embedding a DFT layer at the end of the encoder for Fourier-Net (see Figure~\ref{fig:flowchart1}). Without this layer, such an encoder would be purely a CNN that has to learn complex-valued Fourier coefficients from the spatial image pairs. This setup is similar to DeepFlash \cite{wang2020deepflash}, where two encoders were used to respectively learn the real and imaginary parts of these complex coefficients. As reported in Table \ref{tab:ablation_2doasis_dft}, with this DFT layer the registration Dice score was shown to improve from 0.664 to 0.732 (6.8\%$\uparrow$) and from 0.675 to 0.756 (8.1\%$\uparrow$) for the output sizes of $20\times 24$ and 40$\times$48, respectively. On the other hand, due to the dual encoders, the network required more mult-adds operations and memory footprint to learn the band-limited displacement field. The improvments to both computational efficiency and performance indicate the necessity of using such a DFT layer in our Fourier-Net.

\textbf{Impact of using a band-limited representation:} The necessity of learning a band-limited displacement field might be questioned when one could simply estimate a low-resolution displacement field and then directly up-sample to a full-resolution one using linear interpolation. In Table \ref{tab:ablation_2doasis_band}, we performed such an experiment by replacing the DFT layer and the decoder in Fourier-Net with a simple bilinear interpolation (termed as Bilinear-Net). We observed that in terms of Dice, Bilinear-Net was respectively 1.3\% and 1\% lower than our Fourier-Net for the output size of $20\times 24$ and 40$\times$48. This experiment showed that compared to the low-resolution directly down-sampled displacement field, it was more effective to learn the band-limited representation of the displacement field. As an additional experiment to demonstrate the utility of band-limited images, we evaluated the performance of Fourier-Net+ (see Figure~\ref{fig:flowchart2}) against two variants of Bilinear-Net+: one used bilinear down-sampled images as input and estimated a down-sampled displacement field, and the other one used bilinear down-sampled images as input and estimated a band-limited displacement field. We found that in terms of Dice the two Bilinear-Net+ variants achieved similar results but both are inferior to Fourier-Net+ with a clear performance gap, i.e., 2.2\%$\downarrow$ on resolution $20\times 24$ and 1.2\%$\downarrow$ on resolution 40$\times$48.

\begin{table}[]
\caption{Comparisons between band-limited representations of image and displacements verses up and down-sampling via bilinear-interpolation. 'Input' denotes the input resolution of the $\boldsymbol{CNN}$. $\mathbb{B}_{I_M, I_F}$ indicates the use of the band-limited image or bilinear-downsampled image. $\mathbb{B}_{\boldsymbol{\phi}}$ indicates the use of band-limited displacements or bilinearly-upsampled deformation.}
\label{tab:ablation_2doasis_band}
\centering
\begin{tabular}{lccccc}
\hline
Methods       & $\mathbb{B}_{I_M, I_F}$    & $\mathbb{B}_{\boldsymbol{\phi}}$                   & Input          & Output       & Dice            \\\hline
Bilinear-Net  & --                    & \xmark & $160\times 192$ & $20\times 24$ & 0.719$\pm$0.045 \\
Fourier-Net   & --                    & \cmark & $160\times 192$ & $20\times 24$ & 0.732$\pm$0.042 \\
Bilinear-Net  & --                    & \xmark & $160\times 192$ & $40\times 48$ & 0.746$\pm$0.041 \\
Fourier-Net   & --                    & \cmark & $160\times 192$ & $40\times 48$ & 0.756$\pm$0.039 \\ \hline 
Bilinear-Net+ & \xmark                & \xmark & $40\times 48$   & $40\times 48$ & 0.696$\pm$0.040 \\
Bilinear-Net+ & \xmark                & \cmark & $40\times 48$   & $40\times 48$ & 0.695$\pm$0.040 \\
Fourier-Net+  & \cmark                & \cmark & $40\times 48$   & $40\times 48$ & 0.717$\pm$0.042 \\
Bilinear-Net+ & \xmark                & \xmark & $80\times 96$   & $40\times 48$ & 0.725$\pm$0.040 \\
Bilinear-Net+ & \xmark                & \cmark & $80\times 96$   & $40\times 48$ & 0.726$\pm$0.041 \\
Fourier-Net+  & \cmark                & \cmark & $80\times 96$   & $40\times 48$ & 0.738$\pm$0.041\\\hline
\end{tabular}
\end{table}

\textbf{Diffeomorphisms:} In Table \ref{tab:ablation_2doasis_ss}, we compared the performance of Fourier-Net and Fourier-Net+ and their diffeomorphic counterparts. The squaring and scaling (SS) layers encouraged diffeomorphisms for the estimated deformation, resulting in a lower percentage of negative values of the Jacobian determinant of deformation ($|J|_{< 0}\%$). Besides the influence on negative Jacobians, it was notable that the incorporation of such layers slightly fluctuated the Dice score of different models.
\begin{table}[]
\caption{Comparisions between variants of Fourier-Net and Fourier-Net+, alongside their diffeomorpic counterparts which utilise the squaring and scaling (SS) layer.}
\label{tab:ablation_2doasis_ss}
\centering
\setlength{\tabcolsep}{1.6mm}{
\begin{tabular}{lccccc}
\hline
Methods           & SS                    & Input          & Output       & Dice            & $|J|_{< 0}\%$   \\ \hline
Fourier-Net       & \xmark & $160\times 192$ & $20\times 24$ & 0.732$\pm$0.042 & 0.434$\pm$0.355 \\
Diff-Fourier-Net  & \cmark & $160\times 192$ & $20\times 24$ & 0.735$\pm$0.037 & 0.0$\pm$0.0     \\
Fourier-Net       & \xmark & $160\times 192$ & $40\times 48$ & 0.756$\pm$0.039 & 0.753$\pm$0.408 \\
Diff-Fourier-Net  & \cmark & $160\times 192$ & $40\times 48$ & 0.756$\pm$0.037 & $<$0.0001       \\\hline
Fourier-Net+      & \xmark & $40\times 48$   & $40\times 48$ & 0.717$\pm$0.042 & 0.400$\pm$0.302 \\
Diff-Fourier-Net+ & \cmark & $40\times 48$   & $40\times 48$ & 0.722$\pm$0.038 & 0.0$\pm$0.0     \\
Fourier-Net+      & \xmark & $80\times 96$   & $40\times 48$ & 0.738$\pm$0.041 & 0.674$\pm$0.377 \\
Diff-Fourier-Net+ & \cmark & $80\times 96$   & $40\times 48$ & 0.740$\pm$0.039 & 0.0$\pm$0.0   \\\hline 
\end{tabular}
}
\end{table}

\subsection{Parameter Studies} In this section, we show our investigations into how the choice of different parameter combinations affected the performance of our proposed networks on both 2D OASIS-1 and 3D IXI datasets. 
\begin{table*}[]
\caption{Comparisons between a baseline U-Net backbone and variants of Fourier-Net and Fourier-Net+ on the OASIS-1 and IXI datasets. 'Input' and 'Output' correspond to the input and output resolution of the $\boldsymbol{CNN}$, where the value represents the down-sampling rate on each of the dimensions. For example, $1/2$ will change the resolution to $\frac{H}{2} \times \frac{W}{2} \times \frac{D}{2}$. $\mathbb{B}_{I_M, I_F}$ indicates the use of the band-limited image. We denote the number of cascades in a  cascaded Fourier-Net+ as \textit{K}, and prefix diffeomorphic variants utilizing the squaring and scaling layer with Diff. Multiple metrics such as the \textit{Dice} coefficient, the percentage of negative Jacobian determinant of deformation ($|J|_{< 0}\%$ ), the number of \textit{Mult-Adds} operations in millions (M) or billions (G) required for inference, and the \textit{Memory} footprint in Megabyte (MB) required for one forward/backward pass are reported.}
\label{tab:ablation_2doasis_3DIXI}
\centering

\setlength{\tabcolsep}{1.2mm}{
\begin{tabular}{lccclcccclcccc}
\cline{1-4} \cline{6-9} \cline{11-14}
\multicolumn{1}{c}{\multirow{2}{*}{Methods}} & \multirow{2}{*}{Input} & \multirow{2}{*}{Output} & \multirow{2}{*}{$K$} &  & \multicolumn{4}{c}{OASIS}                                 &  & \multicolumn{4}{c}{IXI}                                      \\ \cline{6-9} \cline{11-14} 
\multicolumn{1}{c}{}                         &                        &                         &                      &  & Dice            & $|J|_{< 0}\%$   & Mult-Adds(M) & Memory &  & Dice            & $|J|_{< 0}\%$   & Mult-Adds (G) & Memory   \\ \cline{1-4} \cline{6-9} \cline{11-14} 
Fourier-Net                                  & 1                      & 1/8                     & --                   &  & 0.732$\pm$0.042 & 0.434$\pm$0.355 & 679.19       & 31.18  &  & 0.754$\pm$0.135 & $<$0.0001       & 135.61        & 4538.28  \\
Diff-Fourier-Net                             & 1                      & 1/8                     & --                   &  & 0.735$\pm$0.037 & 0.0$\pm$0.0     & 679.19       & 31.18  &  & 0.755$\pm$0.131 & 0.0$\pm$0.0     & 135.61        & 4538.28  \\
Fourier-Net                                  & 1                      & 1/4                     & --                   &  & 0.756$\pm$0.039 & 0.753$\pm$0.408 & 888.25       & 35.89  &  & 0.763$\pm$0.129 & 0.024$\pm$0.019 & 169.07        & 4802.93  \\
Diff-Fourier-Net                             & 1                      & 1/4                     & --                   &  & 0.756$\pm$0.037 & $<$0.0001       & 888.25       & 35.89  &  & 0.761$\pm$0.131 & 0.0$\pm$0.0     & 169.07        & 4802.93  \\ \cline{1-4} \cline{6-9} \cline{11-14} 
U-Net                                        & 1                      & 1                       & --                   &  & 0.766$\pm$0.039 & 0.702$\pm$0.343 & 2190         & 96.33  &  & 0.768$\pm$0.127 & 0.134$\pm$0.065 & 868.37        & 16717.97 \\
Diff-U-Net                                   & 1                      & 1                       & --                   &  & 0.762$\pm$0.039 & $<$0.0001       & 2190         & 96.33  &  & 0.765$\pm$0.130 & 0.0$\pm$0.0     & 868.37        & 16717.97 \\ \cline{1-4} \cline{6-9} \cline{11-14} 
Fourier-Net+                                 & 1/4                    & 1/4                     & 1                    &  & 0.717$\pm$0.042 & 0.400$\pm$0.302 & 35.84        & 3.25   &  & 0.736$\pm$0.138 & $<$0.0001       & 3.76          & 142.73   \\
Diff-Fourier-Net+                            & 1/4                    & 1/4                     & 1                    &  & 0.722$\pm$0.038 & 0.0$\pm$0.0     & 35.84        & 3.25   &  & 0.739$\pm$0.134 & 0.0$\pm$0.0     & 3.76          & 142.73   \\
Fourier-Net+                                 & 1/4                    & 1/4                     & 2                    &  & 0.742$\pm$0.038 & 0.370$\pm$0.274 & 71.68        & 6.50   &  & 0.753$\pm$0.135 & 0.0$\pm$0.0     & 7.52          & 285.46   \\
Diff-Fourier-Net+                            & 1/4                    & 1/4                     & 2                    &  & 0.737$\pm$0.038 & 0.0$\pm$0.0     & 71.68        & 6.50   &  & 0.756$\pm$0.130 & 0.0$\pm$0.0     & 7.52          & 285.46   \\
Fourier-Net+                                 & 1/4                    & 1/4                     & 3                    &  & 0.743$\pm$0.038 & 0.282$\pm$0.244 & 107.52       & 9.75   &  & 0.756$\pm$0.133 & $\sim0.0007$    & 11.28         & 428.19   \\
Diff-Fourier-Net+                            & 1/4                    & 1/4                     & 3                    &  & 0.737$\pm$0.037 & $<$0.0001       & 107.52       & 9.75   &  & 0.758$\pm$0.130 & 0.0$\pm$0.0     & 11.28         & 428.19   \\
Fourier-Net+                                 & 1/4                    & 1/4                     & 4                    &  & 0.744$\pm$0.038 & 0.224$\pm$0.207 & 143.36       & 13.00  &  & 0.760$\pm$0.128 & 0.067$\pm$0.042 & 15.04         & 570.92   \\
Diff-Fourier-Net+                            & 1/4                    & 1/4                     & 4                    &  & 0.738$\pm$0.038 & 0.0$\pm$0.0     & 143.36       & 13.00  &  & 0.761$\pm$0.127 & 0.0$\pm$0.0     & 15.04         & 570.92   \\
Fourier-Net+                                 & 1/2                    & 1/4                     & 1                    &  & 0.738$\pm$0.041 & 0.674$\pm$0.377 & 142.66       & 9.52   &  & 0.748$\pm$0.131 & 0.066$\pm$0.051 & 19.30         & 670.20   \\
Diff-Fourier-Net+                            & 1/2                    & 1/4                     & 1                    &  & 0.740$\pm$0.039 & 0.0$\pm$0.0     & 142.66       & 9.52   &  & 0.750$\pm$0.130 & 0.0$\pm$0.0     & 19.30         & 670.20   \\
Fourier-Net+                                 & 1/2                    & 1/4                     & 2                    &  & 0.755$\pm$0.038 & 0.508$\pm$0.274 & 285.32       & 19.04  &  & 0.762$\pm$0.133 & 0.009$\pm$0.008 & 38.60         & 1340.40  \\
Diff-Fourier-Net+                            & 1/2                    & 1/4                     & 2                    &  & 0.749$\pm$0.039 & $<$0.0001       & 285.32       & 19.04  &  & 0.762$\pm$0.130 & 0.0$\pm$0.0     & 38.60         & 1340.40  \\
Fourier-Net+                                 & 1/2                    & 1/4                     & 3                    &  & 0.759$\pm$0.039 & 0.354$\pm$0.278 & 427.98       & 28.56  &  & 0.766$\pm$0.131 & 0.015$\pm$0.011 & 57.90         & 2010.60  \\
Diff-Fourier-Net+                            & 1/2                    & 1/4                     & 3                    &  & 0.753$\pm$0.039 & $<$0.0001       & 427.98       & 28.56  &  & 0.765$\pm$0.128 & 0.0$\pm$0.0     & 57.90         & 2010.60  \\
Fourier-Net+                                 & 1/2                    & 1/4                     & 4                    &  & 0.761$\pm$0.039 & 0.278$\pm$0.232 & 570.64       & 38.08  &  & 0.765$\pm$0.128 & 0.236$\pm$0.115 & 77.204        & 2680.80  \\
Diff-Fourier-Net+                            & 1/2                    & 1/4                     & 4                    &  & 0.755$\pm$0.039 & $<$0.0001       & 570.64       & 38.08  &  & 0.765$\pm$0.128 & 0.0$\pm$0.0     & 77.204        & 2680.80  \\ \hline
\end{tabular}
}
\end{table*}

\textbf{Resolution of band-limited displacement fields:} For the 2D OASIS-1 experiments in Table \ref{tab:ablation_2doasis_3DIXI}, we studied Fourier-Net with two resolutions (i.e., $20\times 24$ and 40$\times$48) of the predicted band-limited displacement field, which were respectively $\frac{1}{8} \times \frac{1}{8} $ and $\frac{1}{4}\times \frac{1}{4} $ of the original image resolution (160$\times$192). It can be seen that resolution $40\times 48$ improved the Dice score by 2.4\% compared to resolution $20\times 24$ (0.732 vs 0.756), with an increase in mult-adds operations (from 679.19M to 888.25M) and memory footprint (from 31.18MB to 35.89MB). Using resolution 40$\times$48, the Dice scores of our Fourier-Net and Diff-Fourier-Net were respectively 1$\%$ and 0.6$\%$ lower than those of the full-resolution U-Net and Diff-U-Net. However, in terms of mult-adds and memory footprint, such two U-Nets were \textbf{2.5} and \textbf{2.7} times more expensive than our Fourier-Net and Diff-Fourier-Net, respectively. For the 3D IXI experiments (with resolutions of $160\times192\times224$ ) in Table \ref{tab:ablation_2doasis_3DIXI}, we also observed that learning a displacement field with a smaller resolution ($20\times 24$$\times$28) was less accurate than using a larger one (40$\times$48$\times$56) in terms of their Dice (0.754 vs 0.763). It also can be seen that learning resolution 40$\times$48$\times$56 performed marginally worse than the full-resolution U-Net and Diff-U-Net, with Dice scores only 0.5$\%$ and 0.4$\%$ lower, respectively. However, in terms of mult-adds and memory footprint, such two U-Nets were \textbf{5.1} and \textbf{3.5} times more expensive than our Fourier-Net and Diff-Fourier-Net, respectively.


\textbf{Resolution of band-limited images:} In Fourier-Net, we selected the optimal resolutions of the band-limited displacements for OASIS ($40\times 48$) and IXI ($40\times 
 48\times 56$). In Fourier-Net+, we also considered the resolution of the band-limited images. In Table \ref{tab:ablation_2doasis_3DIXI}, we experimented on Fourier-Net+ with two resolutions (i.e., $80\times 96$ and 40$\times$48) on 2D OASIS-1. We observed that a larger resolution (80$\times$96) was superior to a smaller one (40$\times$48). Specifically, with $40\times 48$ band-limited images, Fourier-Net+ achieved a Dice score of 0.717, which was 2$\%$ lower than the Fourier-Net+ variant with 80$\times$96. On 3D IXI, Fourier-Net+ achieved a Dice score of 0.748 using resolution 80$\times$96$\times$112, which was 1.2\% higher than that of Fourier-Net+ using resolution  40$\times$48$\times$56. 



\textbf{Impact of cascade number:} As can be seen from Table~\ref{tab:ablation_2doasis_3DIXI}, although Fourier-Net+ significantly reduced the computational cost and memory footprint, its performance was inferior to Fourier-Net. To overcome this accuracy issue, we proposed a cascaded Fourier-Net+ by stacking multiple Fourier-Net+. Given significant computational savings in Fourier-Net+, cascaded Fourier-Net+ still had an efficiency advantage compared to Fourier-Net and U-Net. From Table~\ref{tab:ablation_2doasis_3DIXI}, on OASIS-1, we observe that using more cascades indeed improves performance. For both Fourier-Net+ and Diff-Fourier-Net+ (with resolution $80\times 96$ and $40\times 48$ for input and output respectively), increasing the number of cascades $k$ from 1 to 4 continuously improved the registration performance, i.e., from 0.738 to 0.761 and from 0.740 to 0.755, respectively. Note that, even with 4 cascades, Fourier-Net+ (570.64M) had 35.76\% less mult-adds than Fourier-Net (888.25M), whilst showing performance gains of 0.5$\%$ in terms of Dice (0.761 vs 0.756). Our 4$\times$Fourier-Net+ and Diff-4$\times$Fourier-Net+ were on par with the full-resolution U-Net and Diff-U-Net, with Dice scores only 0.5$\%$ and 0.7$\%$ lower, respectively. However, in terms of mult-adds and memory footprint, such two U-Nets were \textbf{3.8} and \textbf{2.5} times more expensive than our 4$\times$Fourier-Net+ and Diff-4$\times$Fourier-Net+, respectively. We notice a similar improvement of performance with the addition of cascades in 3D IXI: specifically, 3$\times$Fourier-Net+ achieved a 0.766 Dice score which is only 0.2\% lower than that of U-Net. Meanwhile, the Dice score of Diff-3$\times$Fourier-Net+ was the same as Diff-U-Net (0.765), but U-Net and Diff-U-Net had respectively \textbf{15} and \textbf{8.3} times more mult-adds and memory footprint than 3$\times$Fourier-Net+ and Diff-3$\times$Fourier-Net+.
\subsection{Comparison with the state-of-the-art}
We have so far shown that Fourier-Net can learn a band-limited displacement field to represent the full-resolution deformation with minimal performance loss compared to full-resolution U-Net architectures. We then showed, with cascaded Fourier-Net+, that learning such a band-limited displacement field from band-limited images can achieve similar performance with Fourier-Net as well as full-resolution U-Net architectures. In this section, we compare our Fourier-Net and its variants with a few state-of-the-art methods on the three datasets. We note that all reported CPU and GPU runtimes were tested on a machine with 128G RAM, 16 3.80GHz Intel(R) Core(TM) i7-9800X CPUs, and 1 NVIDIA Geforce RTX 2080Ti GPU. The computational time includes the cost of loading models and images and was averaged on the whole testing set with batch size 1.

\subsubsection{Comparison on Inter-subject Brain Registration}

In Table \ref{tab:oasis2d}, we compared the performance of our Fourier-Net and Fourier-Net+ with Flash \cite{zhang2019fast}, DeepFlash \cite{wang2020deepflash}, and Diff-B-Spline \cite{qiu2021learning} on the challenging task of 2D inter-subject brain registration (OASIS-1). We compiled and ran Flash\footnote{\url{https://bitbucket.org/FlashC/flashc/src/master/}} on CPU, but encountered \textit{segmentation fault} errors with the official GPU version. As such we have not compared GPU inference times of Flash. We reported the performance of Flash on three band-limited resolutions (i.e., 16$\times$16, $20\times 24$, and  40$\times$48), and we grid-searched its built-in hyper-parameters over 252 different combinations on the whole validation set for each resolution. We also attempted to run the official DeepFlash\footnote{\url{https://github.com/jw4hv/deepflash}} with supervision from Flash's results. We trained DeepFlash on all 40200 training pairs for up to 1000 epochs with more than 40 different combinations of hyper-parameters. Diff-B-Spline \cite{qiu2021learning} was trained by using its official implementation\footnote{\url{https://github.com/qiuhuaqi/midir}} on all image pairs in the training set. The hyper-parameters were tuned on the held-out validation set, and the highest performing model used MSE data similarity and $\lambda=$0.01 smoothness regularisation.


\begin{table*}[ht]
\caption{Comparing various state-of-the-art methods on the OASIS-1 dataset. The first, second, and third ranked methods in terms of dice score are highlighted in \textcolor{red}{red}, \textcolor{blue}{blue}, and \textcolor{brown}{brown} colors, respectively. We include the patch size of the band-limited displacement/velocity field used by each method, apart from Diff-B-Spline which instead corresponds to the lower dimensional deformation used for cubic B-Spline interpolation.}
\centering
\label{tab:oasis2d}
\begin{small}
\setlength{\tabcolsep}{1.8mm}{
\begin{tabular}{lcccccccc}\hline
Methods     & Resolution & Dice$\uparrow$        & $|J|_{ < 0}\%$  & Parameters   & Mult-Adds (M) & Memory (MB)   & CPU (s) & GPU (s) \\\hline
Initial       & - & 0.544$\pm$0.089 &-           &-&-& - & -      \\
Flash \cite{zhang2019fast}      & 16$\times$16      & 0.702$\pm$0.051 & 0.033$\pm$.126  & - & - & - & 13.699 & - \\
Flash \cite{zhang2019fast}     & $20\times 24$      & 0.727$\pm$0.046 & 0.205$\pm$.279 & - & - & - &  22.575 & -    \\
Flash \cite{zhang2019fast}      & $40\times 48$      & 0.734$\pm$0.045 & 0.049$\pm$.080 & - & -& - & 85.773 & -    \\
DeepFlash  \cite{wang2020deepflash} & 16$\times$16      & 0.615$\pm$0.055 & 0.0$\pm$0.0            & - & - & -& 0.487      &-   \\
DeepFlash  \cite{wang2020deepflash} & $20\times 24$      & 0.597$\pm$0.066            & 0.0$\pm$0.0 & - & - & -&0.617    &-     \\
Diff-B-Spline \cite{qiu2021learning}  & $20\times 24$      & 0.710$\pm$0.041 &  0.014$\pm$0.072 & 70,226 & 86.16 & 6.58   &0.012     &0.015     \\
Diff-B-Spline \cite{qiu2021learning} & $40\times 48$      & 0.737$\pm$0.038  & 0.015$\pm$0.069 & 88,690    & 139.40 & 7.49    &0.012 &0.015 \\\hline
Fourier-Net &$40\times 48$     & \textcolor{brown}{0.756$\pm$0.039}  & 0.753$\pm$0.408       &1,427,376     & 888.25 & 35.89     & 0.011     & 0.015    \\
Diff-Fourier-Net &  $40\times 48$ & \textcolor{blue}{0.756$\pm$0.037}  & $<$0.0001 &1,427,376 & 888.25 & 35.89  & 0.015     & 0.015   \\
Fourier-Net+       & $40\times 48$   & 0.738$\pm$0.041  & 0.674$\pm$0.377 &300,477 & 142.66         & 9.52 & 0.006 & 0.014\\
Diff-Fourier-Net+  & $40\times 48$   & 0.740$\pm$0.039  & 0.0$\pm$0.0    &300,477 & 142.66         & 9.52 & 0.009 & 0.014\\
$4\times$Fourier-Net+ &  $40\times 48$ & \textcolor{red}{0.761$\pm$0.039}  & 0.278$\pm$0.232 &1,201,908 & 570.64        & 38.08 & 0.021 & 0.017\\
Diff-$4\times$Fourier-Net+ &  $40\times 48$ &0.755$\pm$0.039  & $<$0.0001     &1,201,908  & 570.64         & 38.08 & 0.025 & 0.017\\\hline
\end{tabular}

}
\end{small}
\end{table*}
All our variants of Fourier-Net outperformed competing methods in terms of Dice. Compared to Flash using a $40\times 48$ resolution, Diff-Fourier-Net improved Dice by 2.2\% and was \textbf{5718} times faster on CPU. Although DeepFlash was much faster than Flash, we found it extremely difficult to successfully train a model and would expect its true potential to fall close to that of Flash, in line with their published work. We note that DeepFlash is not an end-to-end method, because its output (band-limited velocity field) requires an additional PDE algorithm to compute the final deformation. As such, the method is slower than other deep learning methods such as our methods or Diff-B-Spline (0.012s per image pair on CPU). Fourier-Net+ and Diff-Fourier-Net, with similar mult-adds and memory footprint as Diff-B-Spline, achieved comparable results to Diff-B-Spline in terms of Dice but were respectively 2 and 1.5 times faster than Diff-B-Spline. Our highest performing 4$\times$Fourier-Net+ achieved a Dice score of 0.761, which was able to bridge the gap between Fourier-Net+ and Fourier-Net, whilst still achieving very fast runtimes with fewer mult-adds and less memory footprint. We also listed the percentage of negative values of the Jacobian determinant of deformation (denoted by $|J|_{ < 0}\%$) for all compared methods in Table \ref{tab:oasis2d}. Though both Flash and Diff-B-Spline are diffeomorphic approaches, neither of them produced perfect diffeomorphic deformations on this dataset. Diff-Fourier-Net and Diff-Fourier-Net+ however generated nearly zero negative Jacobian determinants.

\begin{table*}[t!]
\centering
\caption{Performance comparison between different state of the art methods on the 3D IXI dataset. The results of the methods labeled with $\ast$ are taken from TransMorph \cite{chen2021transmorph}, but CPU and GPU runtimes are computed by us on the same machine for a fair comparison. }

\begin{tabular}{lccccccc}
\hline
Methods            & Dice$\uparrow$ & $|J|_{ < 0}\%$     & Parameters & Mult-Adds (G) & Memory & CPU (s) & GPU (s)\\\hline
Affine$^\ast$    & 0.386$\pm$0.195 &- & -                      &-&-&-\\
SyN$^\ast$ \cite{avants_ANTS}       & 0.645$\pm$0.152 & \textless{}0.0001      &-&-&-&-\\
NiftyReg$^\ast$ \cite{modat2010fast} & 0.645$\pm$0.167 & \textless{}0.0001      &-&-&-&-\\
LDDMM$^\ast$ \cite{beg2005computing}     & 0.680$\pm$0.135 & \textless{}0.0001      &-&-&-&-\\
Flash \cite{zhang2019fast}           & 0.692$\pm$0.140 & 0.0$\pm$0.0            &-&-&- & 1760& -\\
deedsBCV$^\ast$ \cite{heinrich2015multi}  & 0.733$\pm$0.126 & 0.147$\pm$0.050        &-&-&-&-\\\hline
VoxelMorph-1$^\ast$ \cite{balakrishnan2019voxelmorph}  & 0.728$\pm$0.129 & 1.590$\pm$0.339    &274,387&304.05&2999.88&2.075&0.398\\
VoxelMorph-2$^\ast$ \cite{balakrishnan2019voxelmorph}  & 0.732$\pm$0.123 & 1.522$\pm$0.336    &301,411&398.81&3892.38&2.321&0.408\\
Diff-VoxelMorph$^\ast$ \cite{balakrishnan2019voxelmorph}  & 0.580$\pm$0.165 & \textless{}0.0001      &307,878&89.67 &1464.26 &1.422 & 0.398\\
Diff-B-Spline$^\ast$ \cite{qiu2021learning}    & 0.742$\pm$0.128 & \textless{}0.0001      &266,387 & 47.05&1233.23 &5.649 & 0.378\\
TransMorph$^\ast$ \cite{chen2021transmorph}       & 0.754$\pm$0.124 & 1.579$\pm$0.328        &46,771,251&657.64 & 4090.31 & 4.094 & 0.516\\
Diff-TransMorph$^\ast$ \cite{chen2021transmorph}  & 0.594$\pm$0.163 & \textless{}0.0001      &46,557,414&252.61& 1033.18 &2.797  &0.419\\
B-Spline-TransMorph$^\ast$ \cite{chen2021transmorph} & 0.761$\pm$0.122 & \textless{}0.0001      &46,806,307&425.95 & 1563.41& 7.582 & 0.417\\
LKU-Net$^\ast$ \cite{jia2022u} & \textcolor{brown}{0.765$\pm$0.129} & 0.109$\pm$0.054       &2,086,342&272.09 & 8713.36& 2.304 & 0.398\\
Diff-LKU-Net$^\ast$ \cite{jia2022u} & 0.760$\pm$0.132 & 0.0$\pm$0.0      &2,086,342&272.09 & 8713.36& 5.914 & 0.390\\\hline
Fourier-Net      & {0.763$\pm$0.129} & 0.024$\pm$0.019       &4,198,352&  169.07 &4802.93& 1.029  &0.384\\
Diff-Fourier-Net & 0.761$\pm$0.131 & 0.0$\pm$0.0     &4,198,352&  169.07 &4802.93&4.668 &0.384\\
Fourier-Net+       & 0.748$\pm$0.131  & 0.066$\pm$0.051 & 880,109& 19.30         & 670.20 & 0.455 & 0.381\\
Diff-Fourier-Net+  & 0.750$\pm$0.130  & 0.0$\pm$0.0     & 880,109& 19.30         & 670.20& 4.035 & 0.378\\
3$\times$Fourier-Net+ & \textcolor{red}{0.766$\pm$0.131}  & 0.015$\pm$0.011 & 2,640,327 & 57.90         & 2010.60 & 2.331 & 0.387 \\
Diff-3$\times$Fourier-Net+ & \textcolor{blue}{0.765$\pm$0.128}  & 0.0$\pm$0.0     & 2,640,327 &57.90         & 2010.60 & 4.981 & 0.396 \\
\hline
\end{tabular}
\label{tab:ixi}
\end{table*}

\begin{figure*}[h!]
  \centering
  \includegraphics[width=0.99\linewidth]{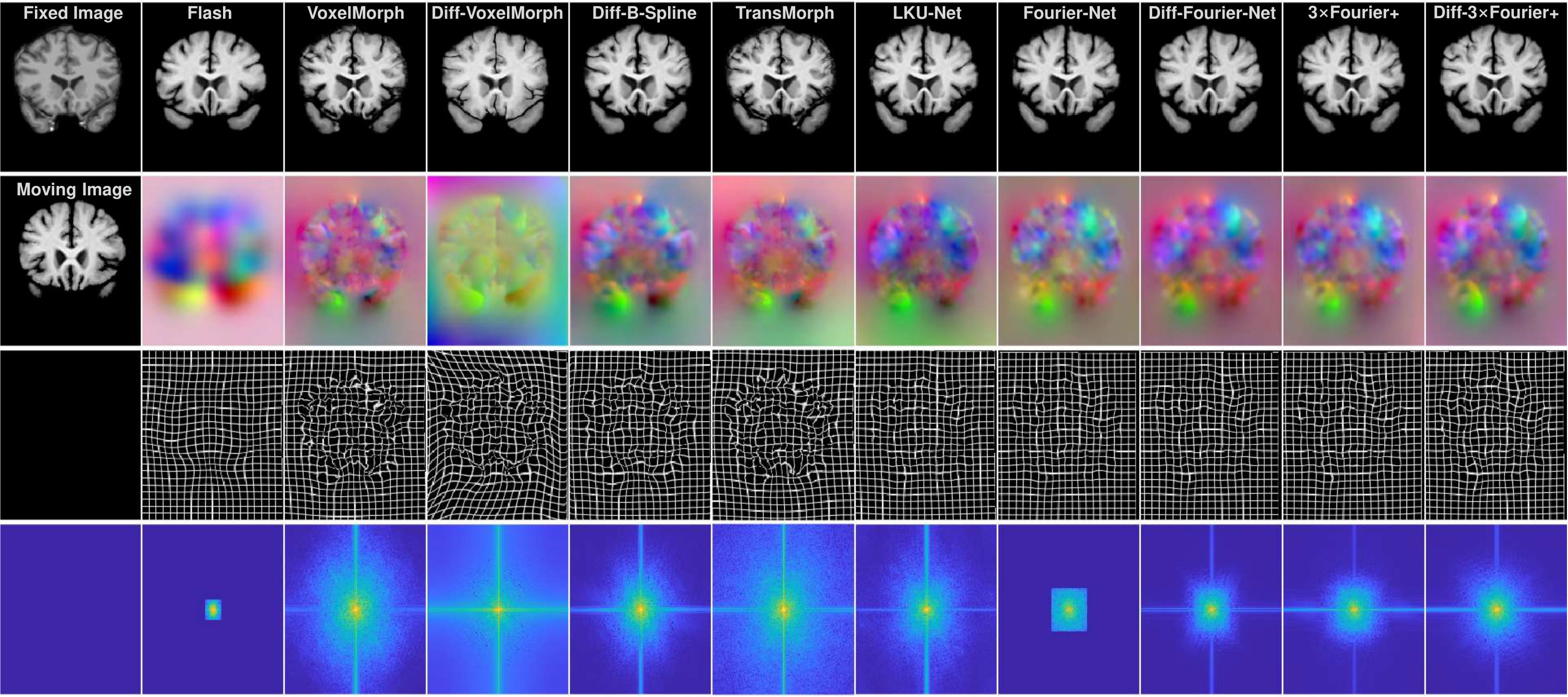}
  \caption{Visual comparison between different methods on the 3D IXI dataset. The 1st column displays a fixed image, a moving image, and two placeholders. From top to bottom, the rows excluding the 1st column are as follows: warped moving images, displacement fields as RGB images, deformation grids, and displacement fields after DFT.}
\label{fig:visualresult}
\end{figure*}

\subsubsection{Comparison on Atlas-Based Brain Registration}

In Table \ref{tab:ixi}, we compared our Fourier-Net and its variants with iterative methods such as ANTs SyN\cite{avants_ANTS} and Flash\cite{zhang2019fast}, as well as deep learning methods such as TransMorph \cite{chen2021transmorph} and LKU-Net \cite{jia2022u}. To guarantee a fair comparison between different methods, we used the IXI dataset with the exact same pre-processing steps and testing protocol as \cite{chen2021transmorph,jia2022u}. Because of this, we directly took relevant results from the original papers \cite{chen2021transmorph} and \cite{jia2022u}, and such results are labeled with $\ast$ in Table \ref{tab:ixi}. Note that the runtimes of all compared methods were computed on our end using the same machine. For Flash \cite{zhang2019fast}, we grid-searched 200 combinations of hyper-parameters using only 5 randomly selected pairs from the validation set, due to the fact that the registration process of Flash takes more than 30 minutes on the CPU for each input image pair.



Our Fourier-Net achieved a 0.763 Dice score which is competitive with top-performing learning-based methods including Transmorph and LKU-Net, whilst reducing CPU inference time to close to a second (1.029s). Fourier-Net+ traded a small performance drop for sub-second CPU runtimes, and had the lowest memory footprint and number of mult-adds across all methods. Specifically, Fourier-Net+ achieved a 0.748 Dice score with only 19.30G mult-adds, 670.2MB memory footprint, and 0.455s per pair speed. Compared to VoxelMorph-1 and VoxelMorph-2 respectively, Fourier-Net+ improved Dice by 2\% and 1.6\% and was \textbf{4.6} and \textbf{4.9} times faster, whilst using only 6.35\% and 4.84\% their mult-adds, and 22.34\% and 17.22\% their memory footprint. Our cascaded 3$\times$Fourier-Net+ and Diff-3$\times$Fourier-Net+ equaled the performance of top state-of-the-art methods whilst retaining a similar inference time compared to those diffeomorphic and non-diffeomorphic learning-based methods. Amongst iterative diffeomorphic methods, Flash achieved the highest Dice score of 0.692, but it needed 1,760s to register an image pair on average. Our Diff-3$\times$Fourier-Net+ achieved a Dice score of 0.765, with 57.90G mult-adds and 4.981s runtimes, outperforming Diff-B-Spline, B-Spline-TransMorph, and Diff-LKU-Net in terms of Dice, mult-adds, and CPU runtimes. We note that those diffeomorphic methods based on dense stationary velocity fields (SVFs) were around 3.6 seconds slower than their non-diffeomorphic versions (see Diff-LKU-Net vs LKU-Net and Diff-Fourier-Net vs Fourier-Net), with the extra 3.6 seconds accounting for the computation of 7 squaring and scaling layers. Such dense SVF-based diffeomorphic methods were however faster than B-Spline methods such as Diff-B-Spline and B-Spline-TransMorph because these methods require additional transposed convolutional layers in order to first recover a full-resolution SVF, which costs extra time. 

\begin{figure*}[h!]
  \centering
  \includegraphics[width=0.99\linewidth]{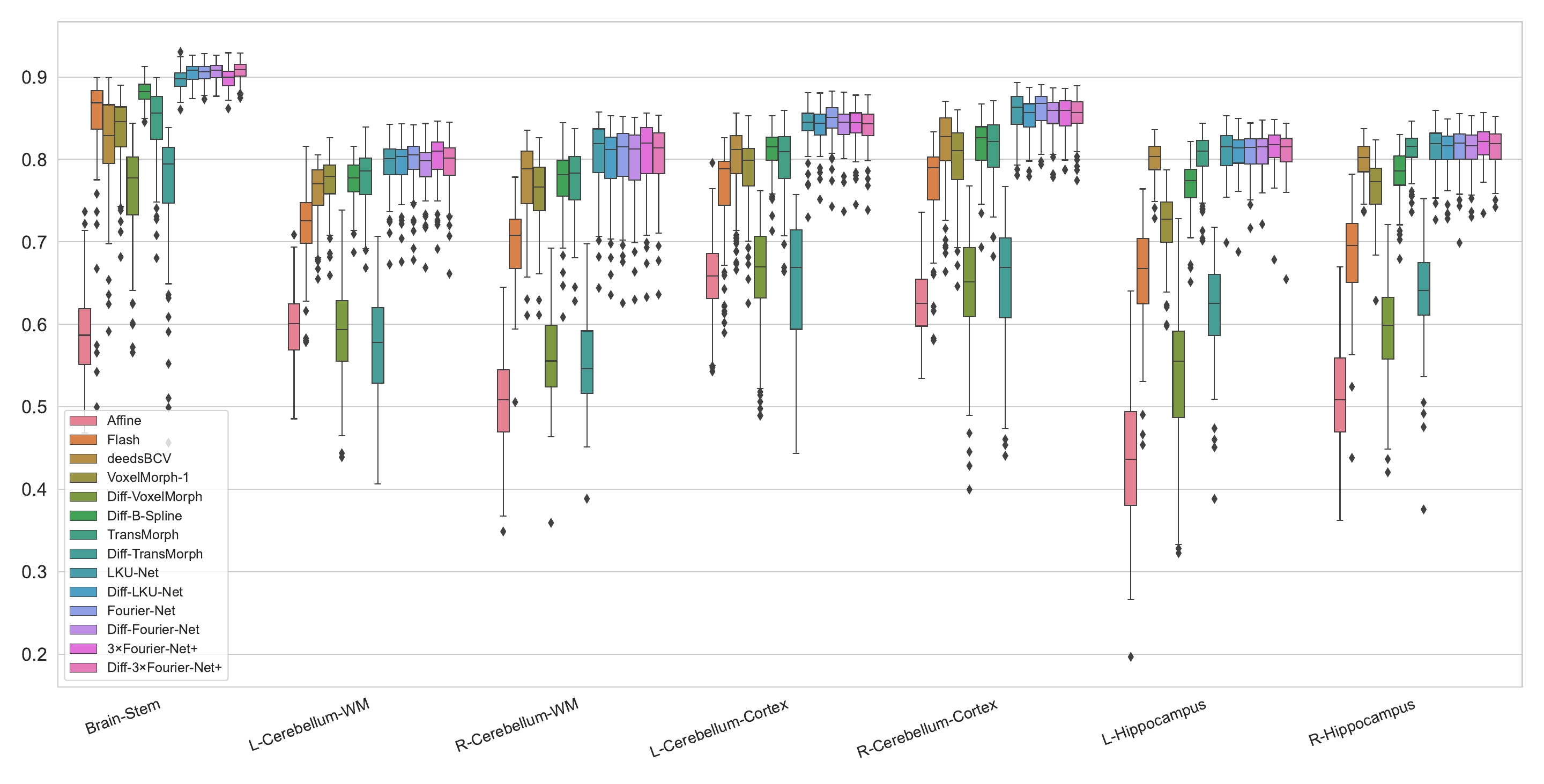}
  \caption{Dice score distributions across the various state-of-the-art methods on different brain structures seen in the 3D IXI dataset, including the brain stem, left/right cerebellum white matter, left/right cerebellum-cortex, and left/right hippocampus.}
\label{fig:brain_boxplot_dice}
\end{figure*}

Figure~\ref{fig:visualresult} shows that whilst Flash's deformation grids have no foldings, it over-smooths the displacement field resulting in a less accurate warping. Figure~\ref{fig:visualresult} (last row) shows that only Flash and Fourier-Net produce strictly band-limited Fourier coefficients. The deformation of Diff-Fourier-Net, 3$\times$Fourier-Net+, and Diff-3$\times$Fourier-Net+ are no longer band-limited due to
the use of the squaring and scaling layers and cascades. We additionally plot the performance of different methods on 7 representative brain structures with the respective boxplot shown in Figure \ref{fig:brain_boxplot_dice}, including the brain stem, left/right cerebellum white matter, left/right cerebellum-cortex, and left/right hippocampus. As can be seen, our Fourier-Net variants consistently perform well over all classes.

\subsubsection{Comparison on 3D Cardiac Motion Estimation}

\begin{table*}[t!]
\centering
\caption{Comparison between different state-of-the-art methods on the 3D-CMR dataset. The top three ranking methods for both Dice and Hausdorff distance (HD) are color-coded \textcolor{red}{red}, \textcolor{blue}{blue}, and \textcolor{brown}{brown} colors,  respectively.}

\begin{tabular}{lcccccccc}
\hline
Methods            & Dice$\uparrow$ & HD$\downarrow$ & $|J|_{ < 0}\%$     & Parameters & Mult-Adds (G) & Memory & CPU (s) & GPU (s)\\\hline
Initial    & 0.493$\pm$0.043 & 8.40$\pm$0.89 &- & -                      &-&-&-\\
Demons       & 0.727$\pm$0.040 & 6.74$\pm$1.00 & 0.0$\pm$0.0      &-&-&-&-\\
FFD       & 0.739$\pm$0.047 & 7.04$\pm$1.20& 0.667$\pm$0.358   &-&-&-&-\\
ANTs SyN  & 0.721$\pm$0.051 & 6.98$\pm$1.23 & 1.388$\pm$0.467        &-&-&-&-\\\hline
VoxelMorph-1 \cite{balakrishnan2019voxelmorph}  & 0.723$\pm$0.032 & 6.53$\pm$1.00& 0.783$\pm$0.325    &274,387&69.50&685.69& 0.459 &0.138  \\
VoxelMorph-2 \cite{balakrishnan2019voxelmorph}  & 0.723$\pm$0.032 & 6.71$\pm$1.04 & 0.695$\pm$0.255    &301,411&91.16&889.69&0.503&0.140\\
3$\times$RC-Net \cite{Zhao_2019_ICCV}    & 0.742$\pm$0.030 & 6.43$\pm$0.90 &0.270$\pm$0.144      &823,161 & 208.5&2057.07 &1.513 & 0.222\\
4$\times$RC-Net \cite{Zhao_2019_ICCV}    & 0.742$\pm$0.030 & 6.46$\pm$0.91 &0.268$\pm$0.145      &1,097,548 & 278.0&2742.76 &2.011 & 0.286\\
SYM-Net\cite{Mok_2020_CVPR}    & 0.731$\pm$0.031 & 6.57$\pm$0.99 &\textless{0.0001}      &1,124,448 & 67.82&1018.31 &1.362 & 0.162\\
VR-Net\cite{jia2021learning}    & 0.733$\pm$0.031 & 6.44$\pm$0.95 & 0.580$\pm$0.185      &825,321 &	211.3	& 2057.07 & 1.388 & 0.231\\
Diff-B-Spline \cite{qiu2021learning}    & 0.809$\pm$0.033 & 6.04$\pm$0.73 & 0.0$\pm$0.0      &211,059 & 5.33&249.71 & 1.457 & 0.137 \\
TransMorph \cite{chen2021transmorph}    & 0.732$\pm$0.029 & 6.51$\pm$0.99 & 0.811$\pm$0.231     &46,771,251 & 150.39&935.16 &1.005 & 0.169\\
LKU-Net \cite{jia2022u}    & 0.728$\pm$0.031 & 6.57$\pm$0.96 & 0.794$\pm$0.282      &2,086,342 & 62.19&1991.62 &0.491 & 0.141 \\
Diff-LKU-Net \cite{jia2022u}    & 0.738$\pm$0.031 & 6.31$\pm$0.94 &\textless{0.0001}      &2,086,342 & 62.19&1991.62 &1.279 & 0.143 \\
\hline
Fourier-Net & 0.814$\pm$0.024  & \textcolor{brown}{6.00$\pm$0.77} & 2.137$\pm$0.658 & 3,891,533 & 31.00 & 1037.32 & 0.225 & 0.140 \\
Diff-Fourier-Net & \textcolor{blue}{0.827$\pm$0.027}  &6.08$\pm$0.85 & $<$0.0003  &3,891,533    & 31.00 & 1037.32 & 0.985 & 0.141\\
Fourier-Net+ & 0.803$\pm$0.031  &6.06$\pm$0.72 & 2.746$\pm$1.033        & 211,081& 0.52 & 19.38 & 0.111 & 0.134\\
Diff-Fourier-Net+ & \textcolor{red}{0.828$\pm$0.030}  &\textcolor{blue}{5.89$\pm$0.81} & \textless{0.0002}        & 211,081& 0.52 & 19.38 & 0.813 & 0.136\\
2$\times$Fourier-Net+ & 0.816$\pm$0.027  &6.02$\pm$0.75 & 1.455$\pm$0.784        & 422,162& 1.04 & 38.76 & 0.245 & 0.135\\
Diff-2$\times$Fourier-Net+ & \textcolor{brown}{0.821$\pm$0.029}  &\textcolor{red}{5.76$\pm$0.70} & 0.0$\pm$0.0        & 422,162& 1.04 & 38.76 & 1.002 & 0.137\\\hline
\end{tabular}
\label{tab:results_3dcmr}
\end{table*}
We conducted a final experiment on the 3D-CMR dataset to ensure we have validated the performance of our models beyond only the task of brain registration. In Table \ref{tab:results_3dcmr}, we compared the registration performance between different methods, including FFD, Demons, ANTs SyN, RC-Net \cite{Zhao_2019_ICCV}, VR-Net\cite{jia2021learning}, TransMorph\cite{chen2021transmorph}, and LKU-Net\cite{jia2022u}. Since we used the same pre-processed data in \cite{jia2021learning}, the results of these iterative methods such as Demons, FFD, and ANTs SyN are directly adopted from \cite{jia2021learning}. We used MSE as the data term for its superior performance to NCC in this dataset. The detailed parameter settings used for
completing deep learning based methods were as follows:
\begin{itemize}
    \item VoxelMorph \cite{balakrishnan2018unsupervised,balakrishnan2019voxelmorph}: We trained two variants of VoxelMorph, i.e., VoxelMorph-1 and VoxelMorph-2. The only difference between them is VoxelMorph-1 has 8 channels at the first and last convolutional layers while VoxelMorph-2 has 16 channels at such layers. Both VoxelMorph-1 and VoxelMorph-2 were trained with the MSE data term with the first-order smoothness regularization, where $\lambda$ was set to 0.01 for optimal results.  
    \item RC-Net \cite{Zhao_2019_ICCV}: We trained two different variants of RC-Net by respectively using 3 and 4 cascades. Both variants used VoxelMorph-1 as the backbone. We therefore term the two variants of RC-Net as 3$\times$RC-Net and 4$\times$RC-Net. The optimal results were achieved with $\lambda$=0.01 for both variants.
    \item SYM-Net \cite{Mok_2020_CVPR}: We used its official code\footnote{\url{https://github.com/cwmok/Fast-Symmetric-Diffeomorphic-Image-Registration-with-Convolutional-Neural-Networks}}. The initial number of kernels was set to 8. The optimal results were achieved with $\lambda$=0.01.
    \item Diff-B-Spline \cite{qiu2021learning}: We trained three different variants of Diff-B-Spline with the control pointing spaces being 3, 4, and 8, respectively. The optimal results were achieved with the control pointing spaces being 8 and $\lambda$ being 0.01.
    \item VR-Net \cite{jia2021learning}: As suggested by the original code\footnote{\url{https://github.com/xi-jia/Learning-a-Model-Driven-Variational-Network-for-Deformable-Image-Registration}}, the data term was set to `$L_1$', the number of warping cascades is set to 2, and the number of intensity consistency layers in each cascade is set to 1 in our experiments. For a fair comparison, we used VoxelMorph-1 as the backbone in each cascade.
    \item TransMorph \cite{chen2021transmorph}: The official TransMorph code\footnote{\url{https://github.com/junyuchen245/TransMorph_Transformer_for_Medical_Image_Registration}} is adopted. The optimal $\lambda$ was set to 0.01.
    \item LKU-Net \cite{jia2022u}: The official implementation\footnote{\url{https://github.com/xi-jia/LKU-Net}} was used, where we set the number of channels in the initial layer as 8 and the large kernel size as 5$\times$5$\times$5 for both LKU-Net and its diffeomorphic version Diff-LKU-Net. The optimal results for both methods were achieved with $\lambda$=0.01.
\end{itemize}

\begin{figure*}[t!]
  \centering
  \includegraphics[width=0.99\linewidth]{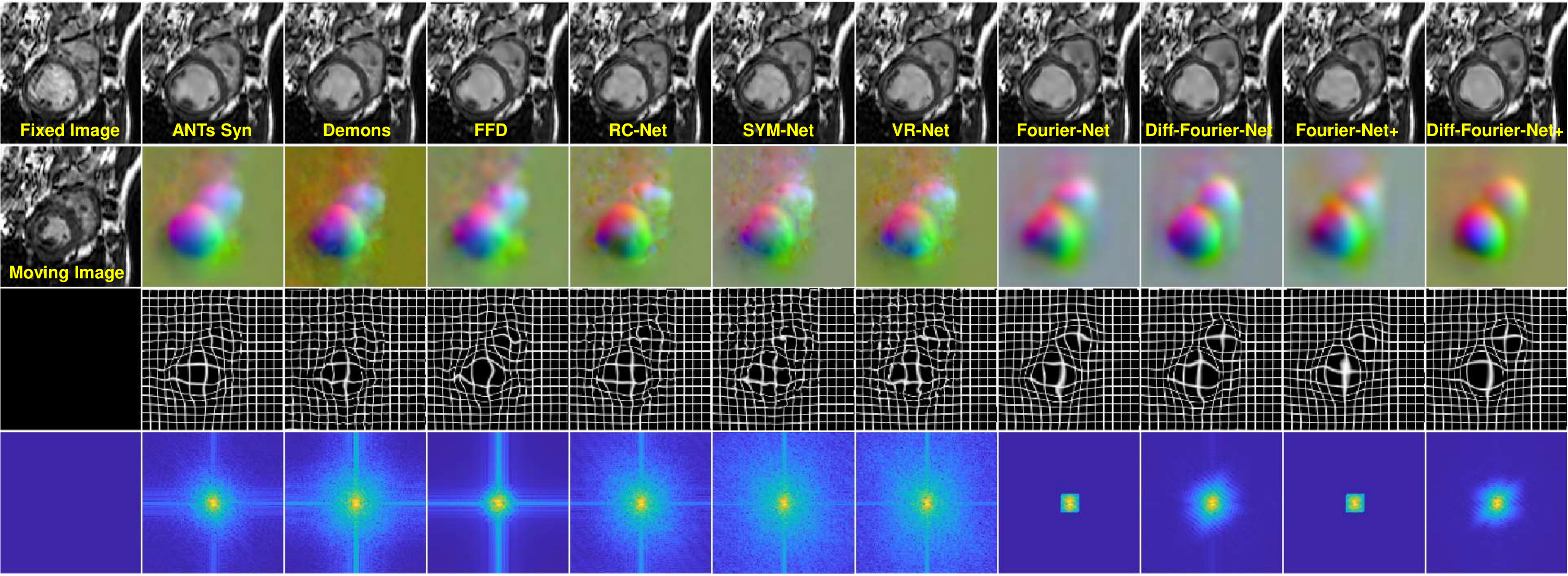}
  \caption{Visual comparison between different methods on 3D-CMR dataset. The 1st column displays a fixed image, a moving image, and two placeholders. From top to bottom, the rows excluding the 1st column correspond to the following: warped moving images, displacement fields as RGB images, deformation grids, and displacement fields after DFT.}
\label{fig:visualresult_3dcmr}
\end{figure*}

\begin{figure*}%
    \centering
    \subfloat[\centering Dice]{{\includegraphics[width=0.49\textwidth]{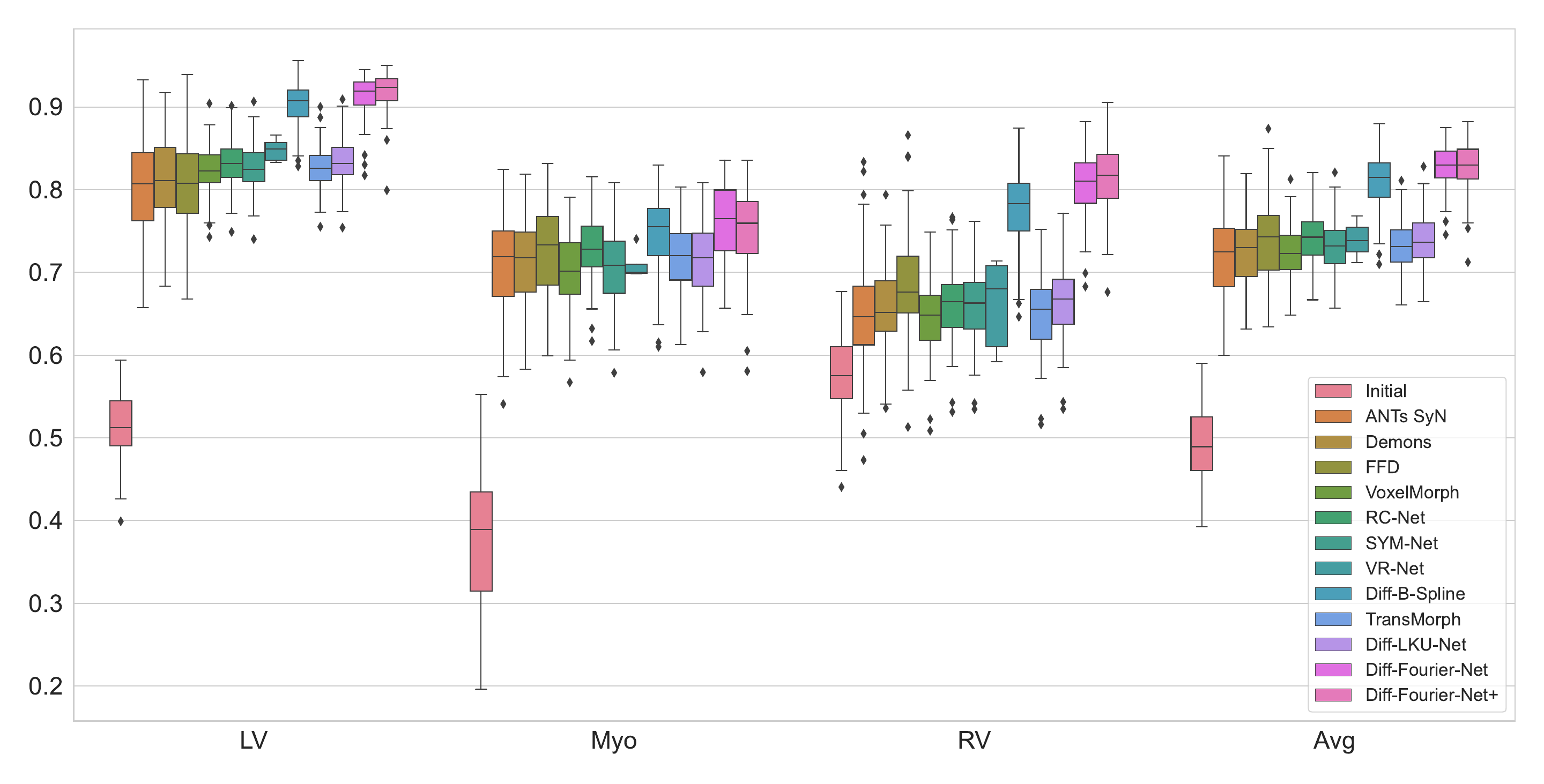} }}%
    \subfloat[\centering Hausdorff Distance]{{\includegraphics[width=0.49\textwidth]{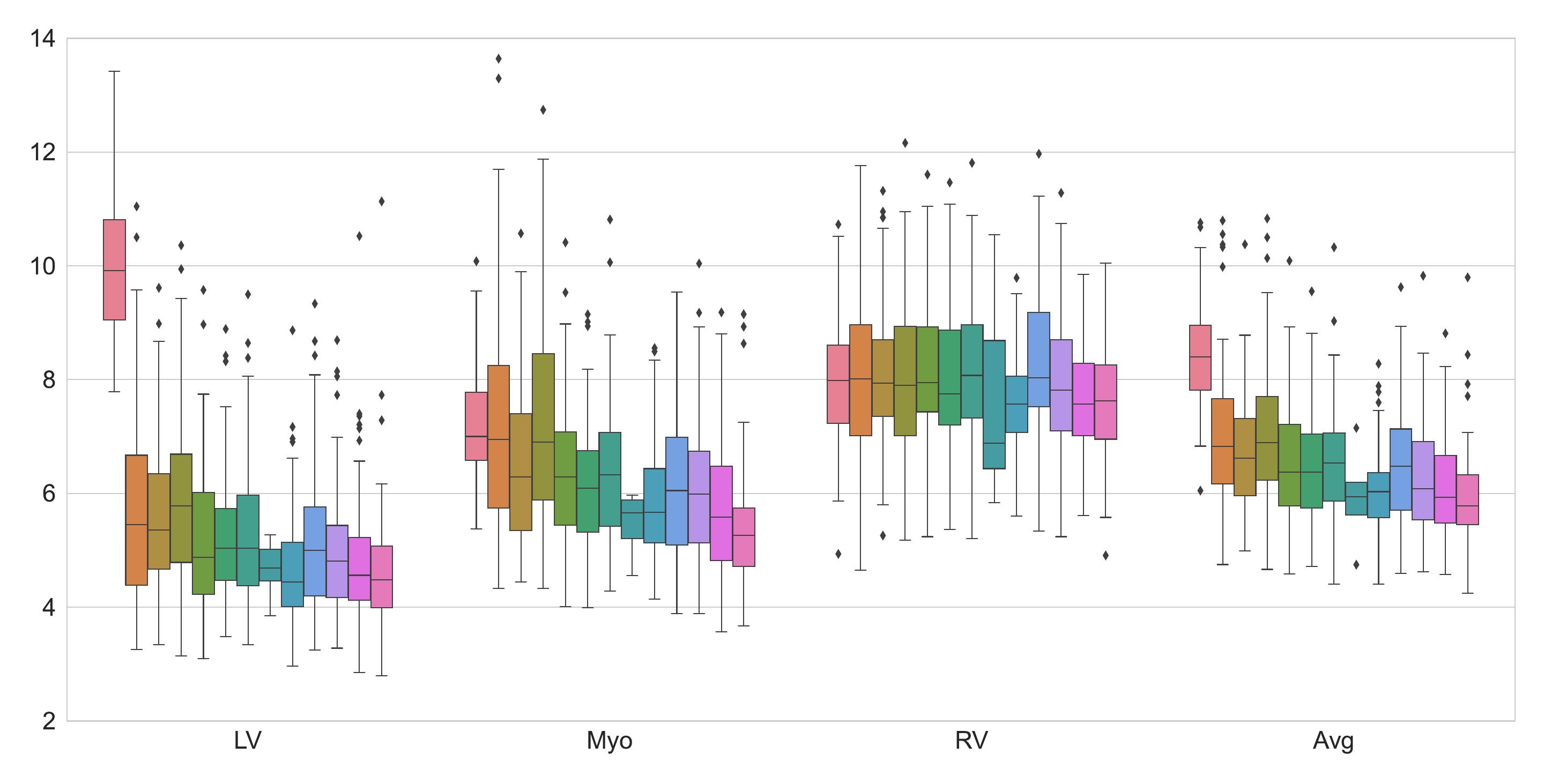} }}%
    \caption{Boxplots depicting the Dice (a) and Hausdorff distance (b) distributions for each state-of-the-art method over the three cardiac structures (LV, Myo, RV) contained in the 3D-CMR dataset, and their average (Avg).}%
    \label{fig:3dcmrboxplot}%
\end{figure*}

\begin{figure*}[t!]
  \centering  \includegraphics[width=0.8\linewidth]{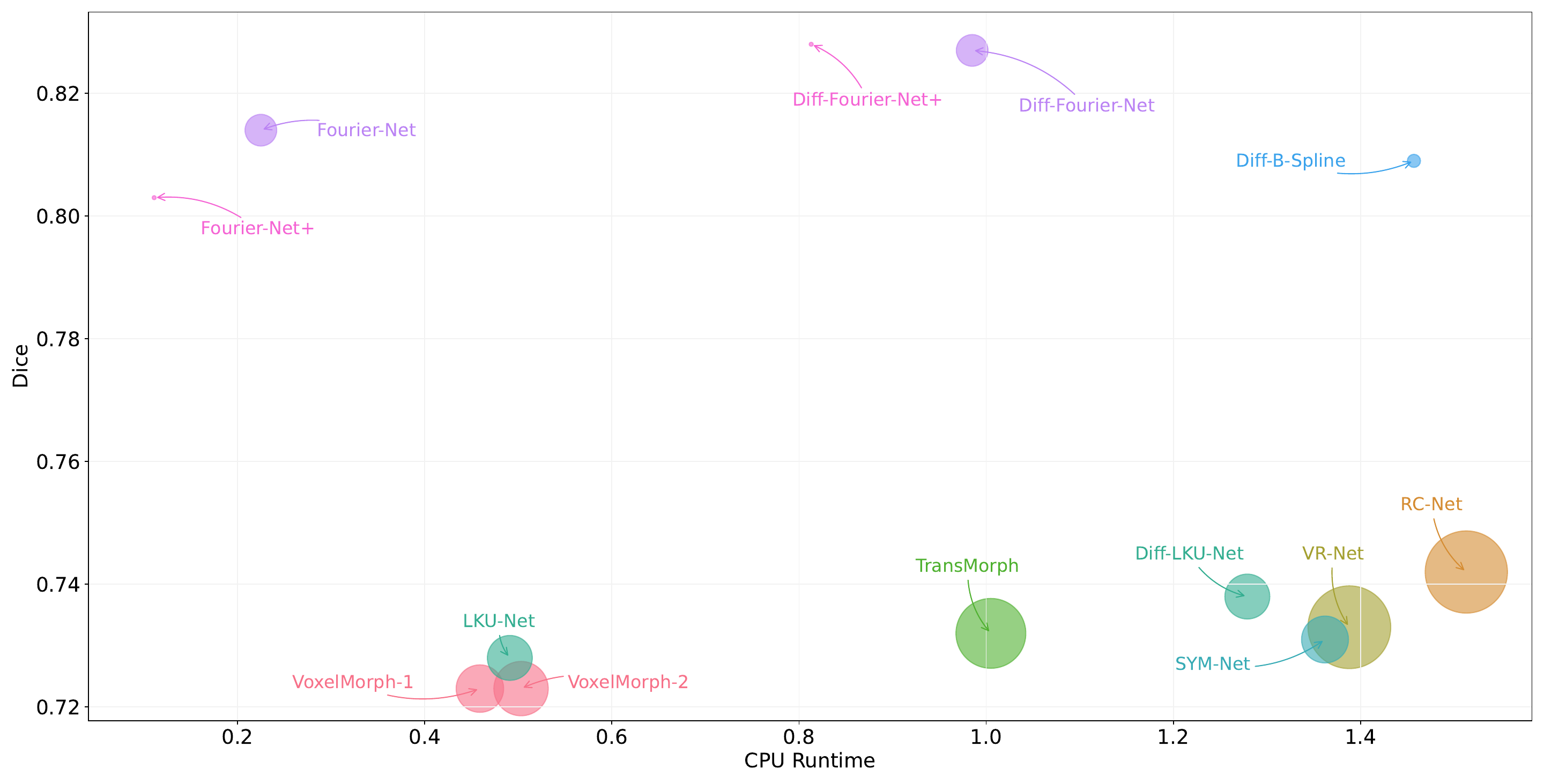}
  \caption{Comparisons of the computational cost of different methods: The $x$-axis and the $y$-axis denote the runtime in CPU (s) and Dice score. The number of mult-adds operations is expressed by the area of the circle.}
\label{fig:3dcmr_ma_dice_time}
\end{figure*}

In Table \ref{tab:results_3dcmr}, we observe that all methods estimating full-resolution deformations including non-diffeomorphic displacement fields and diffeomorphic velocities were outperformed by the methods estimating a low-dimensional representation of the displacement or velocity field, such as Diff-B-Spline and our Fourier-Net variants. We note that the deformations produced by such methods are inherently very smooth. In contrast to the intricate and detailed deformations required in brain registration tasks, the deformation in the left and right ventricles between ED and ES is also smooth. As Dice and Hausdorff distances of the LV, RV and Myo are used as the surrogate for registration accuracy in this task, we believe the inherent smoothness of Diff-B-Spline and Fourier-Nets is an advantage in this task. Although Diff-B-Spline was very competitive on this dataset, the highest Dice and the lowest Hausdorff distance were all achieved by our methods, with our highest performing method (Diff-Fourier-Net+) outperforming Diff-B-Spline by 1.9\% in Dice and 0.15$mm$ in Hausdorff distance (HD). On the other hand, our Fourier-Net+ outperformed TransMorph and LKU-Net, with improvements of 7.1\% and 7.5\% in terms of Dice, and 0.45$mm$ and 0.51$mm$ in terms of HD, respectively. Additionally, our Fourier-Net+ was \textbf{9.05} and \textbf{4.42} times faster while utilizing only 0.35\% and 0.84\% of their multiply-add operations, and 2.07\% and 0.97\% of their memory usage.
 
In Figure. \ref{fig:visualresult_3dcmr}, we plot estimated displacement fields, deformation grids, and warped moving images from each method. We can clearly see that the estimated deformations from our Fourier-Nets are smoother than the completing methods, and Fourier-Net and Fourier-Net+ produce strictly a band-limited deformation. In Figure. \ref{fig:3dcmrboxplot}, we plot the distributions of Dice and HD for different methods over three structures (LV, Myo, and RV) and their average (Avg) on 3D-CMR, where we can clearly observe that our Diff-Fourier-Net and Diff-Fourier-Net+ exhibit an improvement over comparing methods in terms of Dice. The HD distributions of Diff-Fourier-Net are comparable with those of Diff-B-Spline, with a slight improvement seen in Diff-2$\times$Fourier-Net+ on all three structures. In Figure. \ref{fig:3dcmr_ma_dice_time}, we compared the computational cost of different methods, where the $x$-axis and $y$-axis denote the runtime in CPU (s) and Dice score, respectively. The number of mult-adds operations is expressed by the area of a circle. As can be seen Fourier-Net, Fourier-Net+, and cascaded Fourier-Net+ achieved a higher Dice score with faster inference speed and fewer computational costs, while their diffeomorphic versions take a while longer and produce slightly better performance.


\section{Conclusion}
To reduce the computational cost and memory footprint of U-Net style registration networks, we first proposed Fourier-Net to learn the low-dimensional representation of a displacement/velocity field in the band-limited Fourier domain. Building upon Fourier-Net and to further boost the registration efficiency, we then proposed Fourier-Net+ and cascaded Fourier-Net+, aiming to learn the band-limited displacement/velocity field directly from band-limited images, instead of their original full-resolution counterparts. As band-limited images and displacement/velocity fields are of low-resolution representation, our experiments on three datasets showed that Fourier-Net, Fourier-Net+, and cascaded Fourier-Net+ were significantly more efficient than U-Net style architectures and a number of state-of-the-art approaches, whilst retaining a comparative performance in terms of registration accuracy.

Specifically, on the 2D OASIS-1 and 3D IXI brain datasets, we showed that Fourier-Net can learn effectively a band-limited displacement field to represent the full-resolution deformation with minimal performance loss as compared to full-resolution U-Net architectures. We then showed with cascaded Fourier-Net+, that learning such a band-limited displacement field directly from band-limited images was able to achieve similar performance to Fourier-Net and full-resolution U-Net architectures. On the 3D-CMR dataset where cardiac motion is intrinsically smooth and relatively simple, our Fourier-Net+ alone performed already very well, with the registration accuracy on par with Fourier-Net, cascaded Fourier-Net+, and other state-of-the-art methods, but with significantly less computational cost and memory footprint. We also noticed that diffeomorphic Fourier-Net variants were often more accurate than their non-diffeomorphic counterparts in terms of Dice, but were slightly slower in terms of inference speed due to the use of squaring and scaling layers. However, our diffeomorphic methods were still the most efficient approaches when compared to other competing diffeomorphic methods.


Though our proposed Fourier-Net, Fourier-Net+, and cascaded Fourier-Net+ achieved comparable performance with other state-of-the-art methods, it is notable that Fourier-Net assumes that the displacement or velocity field lacks high-frequency signals. This assumption is valid for most smooth and diffeomorphic deformations, and our experimental results on all three datasets also support this assumption. However, we would obviously expect a performance drop in tasks where this assumption does not hold.  In Fourier-Net+, intuitively, one might assume the removal of high-frequency image information within the encoder to be prohibitive in tasks such as brain registration which contain complex structures within images. We showed however that the efficient design of Fourier-Net+ allows the cascaded version of this network to have fewer multiply-add operations than even Fourier-Net, whilst retaining similar performance despite the removal of high-frequency image information in our encoder. We will explore in our future work how to additionally incorporate high-frequency information into Fourier-Net and Fourier-Net+ through the training process to further improve performance.



\section*{Acknowledgments}

The authors would like to Prof Declan P. O’Regan for providing the CMR image data for this research. This work is partially supported by the British Heart Foundation Accelerator Award (AA/18/2/34218), and X. Jia is partially funded by the China Scholarship Council.

 




\section*{Appendix}

\subsection*{3D Cardiac Motion Estimation}

In Table \ref{tab:ablation_3dcmr}, we list more experimental results for this dataset. We can observe that the performance of both 16$\times$16$\times$12 and 32$\times$32$\times$32 Fourier-Net outperforms the full-resolution U-Net backbone with  large margins in terms of Dice, i.e., 7.1\% and 8.3\%. Additionally, the two Diff-Fourier-Nets also outperform the full-resolution Diff-U-Net with  large margins, i.e., 7.9\% and 6.0\%. This phenomenon is different from our observations in the other two brain datasets where Fourier-Net approached to the performance of full-resolution U-Net, but did not exceed it. Here, the results show that through learning a band-limited deformation, Fourier-Net can significantly outperform U-Net. We note that the deformation of the left and right ventricles between ED and ES is very smooth, in contrast to the often intricate and complex deformations required for brain atlas registration. We believe this dataset is therefore particularly well suited to Fourier-Net, as the resultant displacements from the band-limited  $\mathbb{B}_\phi$ intrinsically preserve global smoothness.

In Table \ref{tab:ablation_3dcmr}, we investigate two patch sizes (i.e., 32$\times$32$\times$24 and 64$\times$64$\times$48) for the band-limited image. We observe that even with a 32$\times$32$\times$24 band-limited image as input, the performance  of Fourier-Net+ (0.803 Dice, 6.06$mm$ HD) is already close to Fourier-Net (0.814 and 6.00$mm$). The performance of Diff-Fourier-Net+ (0.828 and 5.89) is also comparable to that of Diff-Fourier-Net (0.827 and 5.89), while the latter has 59.6 times mult-adds and 53.5 times memory footprint. With a 64$\times$64$\times$48 band-limited image, the performance of both Fourier-Net+ and Diff-Fourier-Net+ is further improved.

\begin{table*}[!ht]
\caption{Parameter studies on the 3D CMR dataset. The Dice score and Hausdorff Distance (HD) are computed by averaging that of LV, Myo, and RV of all subjects in the test set.}
\centering
\label{tab:ablation_3dcmr}
\begin{tabular}{lcccccccccc}
\hline
Methods                 & Input                     & Output                    & $K$    & Dice             & HD           & $|J|_{< 0}\%$   & Mult-Adds (G) & Memory (MB)\\\hline
Fourier-Net             & 128$\times$128$\times$96  &16$\times$16$\times$12     & --     & 0.814$\pm$0.024  &6.00$\pm$0.77 & 2.137$\pm$0.658 & 31.00 & 1037.32\\
Diff-Fourier-Net        & 128$\times$128$\times$96  &16$\times$16$\times$12     & --     & 0.827$\pm$0.027  &6.08$\pm$0.85 & $<$0.0003     & 31.00 & 1037.32\\
Fourier-Net             & 128$\times$128$\times$96  & 32$\times$32$\times$24    & --     & 0.813$\pm$0.025  &6.74$\pm$0.85 & 3.801$\pm$0.862 & 38.65 & 1097.81\\
Diff-Fourier-Net        & 128$\times$128$\times$96  & 32$\times$32$\times$24    & --     & 0.813$\pm$0.029  &6.15$\pm$0.81 & 0.001$\pm$0.002 & 38.65 & 1097.81\\\hline
U-Net                   & 128$\times$128$\times$96  & 128$\times$128$\times$96  & --     & 0.730$\pm$0.030  &6.45$\pm$0.95 & 0.726$\pm$0.275        & 198.49 & 3821.25\\
Diff-U-Net              & 128$\times$128$\times$96  & 128$\times$128$\times$96  & --     & 0.742$\pm$0.031  &6.24$\pm$0.92 & $<$0.0001        & 198.49 & 3821.25\\\hline
Fourier-Net+            & 32$\times$32$\times$24    & 16$\times$16$\times$12    & 1      & 0.803$\pm$0.031  &6.06$\pm$0.72 & 2.746$\pm$1.033        & 0.52 & 19.38\\
Diff-Fourier-Net+       & 32$\times$32$\times$24    & 16$\times$16$\times$12    & 1      & 0.828$\pm$0.030  &5.89$\pm$0.81 & \textless{0.0002}        & 0.52 & 19.38 \\
Fourier-Net+            & 32$\times$32$\times$24    & 16$\times$16$\times$12    & 2      & 0.816$\pm$0.027  &6.02$\pm$0.75 & 1.455$\pm$0.784        & 1.04 & 38.76 \\
Diff-Fourier-Net+       & 32$\times$32$\times$24    & 16$\times$16$\times$12    & 2      & 0.821$\pm$0.029  &5.76$\pm$0.70 & 0.0$\pm$0.0        & 1.04 & 38.76 \\
Fourier-Net+            & 32$\times$32$\times$24    & 16$\times$16$\times$12    & 3      & 0.819$\pm$0.032  &5.87$\pm$0.75 &  1.203$\pm$0.675        & 1.56 & 58.14 \\
Diff-Fourier-Net+       & 32$\times$32$\times$24    & 16$\times$16$\times$12    & 3      & 0.823$\pm$0.030  &5.82$\pm$0.82 &  $<$0.0001        & 1.56 & 58.14 \\
Fourier-Net+            & 64$\times$64$\times$48    & 16$\times$16$\times$12    & 1      & 0.811$\pm$0.028  &6.03$\pm$0.73 & 2.145$\pm$0.888        & 3.85 & 134.70\\
Diff-Fourier-Net+       & 64$\times$64$\times$48    & 16$\times$16$\times$12    & 1      & 0.829$\pm$0.026  &5.87$\pm$0.76 & 0.013$\pm$0.073        & 3.85 & 134.70\\
Fourier-Net+            & 64$\times$64$\times$48    & 16$\times$16$\times$12    & 2      & 0.821$\pm$0.027  &5.83$\pm$0.85 & 1.600$\pm$0.775         & 7.70 & 269.40\\
Diff-Fourier-Net+       & 64$\times$64$\times$48    & 16$\times$16$\times$12    & 2      & 0.830$\pm$0.028  &5.69$\pm$0.70 & $<$0.0003         & 7.70 & 269.40\\
Fourier-Net+            & 64$\times$64$\times$48    & 16$\times$16$\times$12    & 3      & 0.820$\pm$0.023  &6.12$\pm$0.72 & 0.560$\pm$0.412         & 11.55 & 404.10\\
Diff-Fourier-Net+       & 64$\times$64$\times$48    & 16$\times$16$\times$12    & 3      & 0.825$\pm$0.025  &5.74$\pm$0.74 & \textless{0.0002}         & 11.55 & 404.10\\
\hline
\end{tabular}
\end{table*}

\subsection*{Hyper-parameters of Flash and DeepFlash}
In this section, we detail the hyper-parameters of Flash and DeepFlash used for our experiments in the main paper.
\subsubsection*{Flash on 2D OASIS}
In Table 2 of the main paper, we reported the registration performance of Flash \cite{zhang2019fast} on 2D OASIS data with respect to three patch sizes of band-limited velocities, i.e., 16$\times$16, 20$\times$24, and 40$\times$48. The 16$\times$16 patch is recommended in the official implementation. However, For a fair comparison with our Fourier-Net, we additionally experimented the patch sizes of 20$\times$24 and 40$\times$48 as well.

For each patch size, by varying $\alpha$ ($\alpha \in \{1,3,4,6\}$), $\gamma$ ($\gamma \in \{0.05,0.1,0.2,0.5,1.0, 2.0,5.0\}$), and $\sigma$ ($\sigma \in \{0.01,0.03,0.05,0.1,0.3,0.5,1,3,5\}$) appeared in the official implementation\footnote{\url{https://bitbucket.org/FlashC/flashc/src/master/Testing/runImageMatching/runImageMatching.sh}}, we experimented 252 different hyper-parameter combinations. We then used the set of hyper-parameters that has the best performance on the validation set and reported its registration performance on the test set. In Table \ref{tab:2Dflash}, we list the final hyper-parameters we used for the 2D OASIS test set.
\begin{table}[ht]
\caption{Hyper-parameters of Flash used for 2D OASIS test dataset.}
\centering

\resizebox{\columnwidth}{!}{
\label{tab:2Dflash}
\begin{tabular}{cccccccc}
\hline
Patch Size & $\sigma$ & $\alpha$ & $\gamma$ & numStep & lpower & stepSizeGD & maxIter \\
\hline
$16\times16$   &0.5    &3.0     &0.2     & 10      & 6      & 0.05       & 200     \\
$20\times24$   &0.05    &3.0     &0.5     & 10      & 6      & 0.05       & 200     \\
$40\times48$   &0.05    &3.0     &0.5     & 10      & 6      & 0.05       & 200 \\
\hline  
\end{tabular}}

\end{table}

\subsubsection*{Flash on 3D IXI}

In Table 4 of the main paper, we reported the performance of Flash on the 3D IXI test set. For this experiment, we tried in total 200 different combinations of hyper-parameters with three different patch sizes ($16\times16\times16$,$20\times24\times28$, and $40\times48\times56$). The experiments were performed on 5 randomly selected validation examples only, as Flash takes really long time to predict a band-limited velocity field for an image pair (30 minutes for $20\times24\times28$ band-limited velocity fields and more than 1 hour for $40\times48\times56$ band-limited velocity fields). In Table \ref{tab:3Dflash}, we listed the final hyper-parameters used in the test set. As the best result came from using the $20\times24\times28$ patch size, the Fourier coefficients of Flash in Figure 4 of the main paper have the smallest area.
\begin{table}[t]
\caption{Hyper-parameters of Flash used for 3D IXI test dataset.}
\centering

\resizebox{\columnwidth}{!}{
\label{tab:3Dflash}
\begin{tabular}{cccccccc}
\hline
Patch Size & $\sigma$ & $\alpha$ & $\gamma$ & numStep & lpower & stepSizeGD & maxIter \\
\hline
$20\times24\times28$   &0.2    &6.0     &0.3     & 10      & 6      & 0.02       & 200     \\
\hline  
\end{tabular}}
\end{table}
\subsubsection*{DeepFlash on 2D OASIS}
In Table 2 of the main paper, we reported the registration performance of DeepFlash \cite{wang2020deepflash} on 2D OASIS data with two different patch sizes, i.e., 16$\times$16 and 20$\times$24. We did not implement DeepFlash with 40$\times$48 patch size because: 1) Flash takes 85.773 seconds to predict 40$\times$48 band-limited velocity fields, producing the velocity fields for all 40200 training pairs therefore will cost about 40 days, which is not practical; and 2) the performance of DeepFlash is bounded by Flash, whose performance in terms of Dice is already lower than that of the proposed Fourier-Net. For the similar reasons, we did not include DeepFlash on 3D IXI either.

We experimented 40 combinations of hyper-parameters by varying learning rate ($\{0.005, 0.001, 0.0005, 0.0001\}$), batch size ($\{1,8,16,32,64\}$) and dropout ratio ($\{0, 0.2\}$) in DeepFlash. The maximum training epoch for each set of parameters is 1000 epochs, as we notice that DeepFlash is slow to converge. Note that we only train our Fourier-Net with 10 epochs on this 2D data.
The final hyper-parameters of training DeepFlash are listed in Table \ref{tab:2Ddeepflash}.

\begin{table}[h!]

\caption{Hyper-parameters used to train DeepFlash.}
\centering

\resizebox{\columnwidth}{!}{
\label{tab:2Ddeepflash}
\begin{tabular}{cccccccc}
\hline
Patch Size & Batch Size & Learning Rate & Dropout Ratio \\
\hline
$16\times16$   &64    &0.0001     &0.2     \\
$20\times24$   &64    &0.0001     &0.2     \\
\hline  
\end{tabular}}
\end{table}

\bibliographystyle{IEEEtran}
\bibliography{IEEEabrv,ref}
 
\end{document}